# Finding Maximum Weight 2-Packing Sets on Arbitrary Graphs

**Jannick Borowitz** | **Ernestine Großmann** | **Christian Schulz**

[1]Faculty of Mathematics and Informatics, University Heidelberg, Baden-Wurtemberg, Germany

**Correspondence**
Ernestine Großmann,
Email: e.grossmann@informatik.uni-heidelberg.de

Funding Information
This research was supported by the DFG grant SCHU 2567/3-1.

**Abstract**

A 2-packing set for an undirected, weighted graph $G = (V, E, w)$ is a subset $\mathcal{S} \subseteq V$ such that any two vertices $v_1, v_2 \in \mathcal{S}$ are not adjacent and have no common neighbors. The Maximum Weight 2-Packing Set problem that asks for a 2-packing set of maximum weight is **NP**-hard. Next to 13 novel data reduction rules for this problem, we develop two new approaches to solve this problem on arbitrary graphs. First, we introduce a preprocessing routine that exploits the close relation of 2-packing sets to independent sets. This makes well-studied independent set solvers usable for the Maximum Weight 2-Packing Set problem. Second, we propose an iterative reduce-and-peel approach that utilizes the new data reductions. Our experiments show that our preprocessing routine gives speedups of multiple orders of magnitude, while also improving solution quality, and memory consumption compared to a naive transformation to independent set instances. Furthermore, it solves 44 % of the instances tested to optimality. Our heuristic can keep up with the best-performing maximum weight independent set solvers combined with our preprocessing routine. Additionally, our heuristic can find the best solution quality on the biggest instances in our data set, outperforming all other approaches.

## 1 | INTRODUCTION

For a given undirected graph $G = (V, E)$ a *2-packing set* is defined as a subset $\mathcal{S} \subseteq V$ of all vertices such that for each pair of distinct vertices $v_1 \neq v_2 \in \mathcal{S}$ the shortest path between $v_1$ and $v_2$ has at least three edges. A *maximum 2-packing set* (M2PS) is a 2-packing set with highest cardinality. For a weighted graph $G = (V, E, w)$ with non-negative vertex weights given by a function $w : V \to \mathbb{R}_{\geq 0}$, the problem is generalized to its weighted version the Maximum Weight 2-Packing Set (MW2PS) problem. The objective here is to find a 2-packing set $\mathcal{S}$ of maximum weight, i.e. such that $w(\mathcal{S}) = \sum_{v \in \mathcal{S}} w(v)$ is maximum. The Maximum (Weight) $k$-Packing Set problem is a further generalization, where the shortest path contains at least $k + 1$ edges. Sometimes, it is also referred to as the problem of finding a maximum (weight) distance-$d$ independent set where $d = k + 1$. For $k = 1$, this results in the Maximum (Weight) Independent Set (M(W)IS) problem.

An important application for the Maximum 2-Packing Set problem is given in distributed algorithms. In contrast to the Maximum Independent Set problem, which, given a solution vertex, only conflicts with the direct neighborhood, the 2-packing set provides information about a larger area around the vertex. This is important for self-stabilizing algorithms [22, 23, 41, 46, 51, 53, 54]. In particular, computing large 2-packing sets can be used as a subroutine to ensure mutual exclusion of vertices with overlapping neighborhoods. An example is finding a minimal $\{k\}$-dominating function [23] which has various applications itself as presented by Gairing *et al.*[21]. Bacciu *et al.*[5] use large $k$-packing sets to develop a Downsampling approach for graph data. This is particularly useful for deep neural networks. Further, Soto *et al.*[30] show that the knowledge of the size of a maximum 2-packing set in special graphs can be used for error correcting codes, and Hale *et al.*[31] indicate that large 2-packing sets can be used to model interference issues for frequency assignment. This can be done by looking at the frequency-constrained co-channel assignment problem. In this application, the vertex set consists of locations of radio transmitters, and two vertices share an edge if the frequencies are mutually perceptible. We want to assign a channel to as many radio transmitters as possible to conserve spectrum and avoid interference. Therefore, vertices assigned to the same channel must have a certain distance. A weighted extension to this application is to assign different importance to the locations modeled by the vertex weights. The MW2PS is a set of most important locations for radio transmitters, with a distance of three, to avoid interference. The M2PS problem is also of theoretical interest since it is used to compute neighborhood-restricted [≤2]-achromatic colorings [10] and the Roman domination function [13].

As an **NP**-hard problem [33], the running time of an exact solver for the MW2PS problem grows exponentially with the size of the graph. *Data reduction rules* are a powerful technique for tackling **NP**-hard graph problems. These are polynomial time procedures that remove or contract local graph structures. After applying a sequence of reduction rules, the input instance is reduced to an equivalent, smaller instance. Originally developed as a tool for parameterized algorithms [15], data reduction rules have been effective in practice for computing a Maximum (Weight) Independent Set [12, 28, 36, 47, 26] / Minimum Vertex Cover [3, 38], Maximum Clique [11, 55], and Maximum $k$-Plex [14, 34], as well as solving Graph Coloring [40, 55] and clique cover problems [25, 48]. For further details on data reduction rules, we direct the reader to the survey by Abu-Khzam *et al.* [2].

Borowitz *et al.* [9] where the first to investigate data reduction rules for the Maximum 2-Packing Set problem. The MW2PS problem and M2PS problem can be solved with well-studied independent set solvers by using a graph transformation. This transformation expands the 2-neighborhood of each vertex and builds the square graph. Much engineering has yielded highly scalable solvers for the MWIS problem in recent years. However, graphs can become very dense if we directly compute the square graph, prohibiting scalability. Thus, we present new data reductions specific to the MW2PS problem, first applied to the original problem instance. Afterward, we compute the square graph on which MWIS solvers can be applied more efficiently.

*Our Results.* We introduce the first 13 data reduction rules for the MW2PS problem, which we utilize in our preprocessing routine reduce&transform. This routine transforms an instance of the MW2PS problem into an instance of the MWIS problem, such that the solution to the MWIS problem on the transformed graph is an MW2PS on the original graph. With this routine, scalable MWIS solvers can solve the MW2PS problem efficiently. Our experiments show that this routine can fully reduce and thereby optimally solve 44 % of the instances in our data set. Furthermore, it improves solution quality, running time, and memory consumption compared to the square graph transformation.

Moreover, we propose a new heuristic based on the metaheuristic Concurrent Difference Core Heuristic introduced by Großmann *et al.* [26]. Our heuristic excels especially for large graphs, where the graph transformation uses much memory. Our experiments indicate that our heuristic can keep up with the state-of-the-art MWIS solvers equipped with our preprocessing routine. Additionally, it can find the best solution quality on the biggest instances in our data set, outperforming all MWIS approaches.

## 2 | PRELIMINARIES

Let $G = (V, E, w)$ be an undirected, vertex-weighted graph with $n = |V|$ and $m = |E|$, where $V = \{0, \ldots, n-1\}$. The function $w : V \to \mathbb{R}_{\geq 0}$ assigns each vertex a non-negative weight. This function is extended to a set of vertices $U \subset V$ by $w(U) = \sum_{v \in U} w(v)$. Additionally, we define the abbreviation $w_{max}(U) = \max\{w(u) \mid u \in U\}$. A path $p$ in $G$ is a sequence of adjacent edges. The *length* of the path is equal to its number of edges. We extend the graph definition to a *link-graph* $\mathcal{G} = (G, \mathcal{L})$, which is a tuple of a graph $G$ and a set of *links* $\mathcal{L} \subseteq \binom{V}{2}$. This link set $\mathcal{L}$ is disjunct to the edge set $E$, i.e. $\mathcal{L} \cap E = \emptyset$. Two vertices connected by a link are called *linked* vertices. In the link-graph $\mathcal{G}$, a path can also contain links. For each link in the path, we add 2 to its length. Therefore, the shortest path between two linked vertices in $\mathcal{G}$ is of length 2. The *induced subgraph* of a graph $G = (V, E)$ of a set of vertices $U \subseteq V$ is defined as $G[U] = (U, E[U])$ with $E[U] = \{\{u, v\} \in E \mid u, v \in U\}$. Similarly, we define the *induced link-subgraph* of a set of vertices $U \subseteq V$ as $\mathcal{G}[U] = (G[U], \mathcal{L}[U])$ with $\mathcal{L}[U] = \{\{u, v\} \in \mathcal{L} \mid u, v \in U\}$. We use the short notation $\mathcal{G} - v$ for $\mathcal{G}[V \setminus \{v\}]$ and $\mathcal{G} - U$ for $\mathcal{G}[V \setminus U]$. The open neighborhood $N(v)$ of a vertex $v \in V$ is defined as $N(v) = \{u \in V \mid \{u, v\} \in E\}$ and the closed neighborhood $N[v] = N(v) \cup \{v\}$. The notation is extended to a set of vertices $U \subseteq V$ with $N(U) = \cup_{u \in U} N(u) \setminus U$ and $N[U] = N(U) \cup U$. Similarly, we define the *link-neighborhood* of a vertex $v \in V$ as $L(v) = \{u \in V \mid \{u, v\} \in \mathcal{L} \vee u \in N(N(v))\}$. The closed 2-neighborhood is defined as $N_2[v] = N[v] \mathbin{\dot{\cup}} L(v)$ and the open 2-neighborhood by $N_2(v) = N_2[v] \setminus \{v\}$. By this definition, for all vertices $u \in L(v)$, the shortest path from $u$ to $v$ is of length 2. The degree of a vertex $v$ is the size of its neighborhood $\deg(v) = |N(v)|$. The *link-degree* of a vertex is defined by the size of its link-neighborhood $\deg_{\mathrm{L}}(v) = |L(v)|$. We define the maximum degree as $\Delta = \max_{v \in V} \deg(v)$. The *square graph* $G^2 = (V, E^2, w)$ of a graph $G = (V, E, w)$ is defined as a graph with the same vertex set and weights, but an extended edge set $E^2 = E \cup \tilde{E}$. For every pair of non-adjacent vertices $u, v \in V$ that share a common neighbor in $G$, we have $\{u, v\} \in \tilde{E}$. For $0 < k \in \mathbb{N}$ a *k-packing set* is defined as a subset $\mathcal{S} \subseteq V$ such that between each pair of vertices in $\mathcal{S}$ the shortest path has length at least $k + 1$. For $k = 1$, we refer to the set as the *independent set*, where all vertices in $\mathcal{S}$ are non-adjacent. The *maximum* independent set (MIS) problem is finding an independent set of maximum cardinality. This work mainly focuses on the weighted generalization of the case $k = 2$, yielding a weighted 2-packing set. In this set, the shortest path between each pair of vertices is at least length three. A *maximal* 2-packing set is a 2-packing set $\mathcal{S} \subseteq V$ that cannot be extended by any further vertex $v \in V$ without violating the 2-packing set condition. The *maximum* 2-packing set problem (M2S) is finding a 2-packing set of maximum cardinality. The Maximum Weight 2-Packing Set (MW2PS) problem is to find a 2-packing set of maximum weight. Analogously to the independence number $\alpha(G)$ for the Maximum Independent Set problem, we define $\alpha_w^2(\mathcal{G})$ as the weight of the solution to the Maximum 2-Packing Set problem for a link-graph $\mathcal{G}$. For a graph $G$, we define $\alpha_w^2(G) = \alpha_w^2((G, \emptyset))$.

We define a *distance-2-clique* as a set of vertices in $\mathcal{G}$ whose vertices are pairwise connected by a path of length at most 2. A vertex $v$ is called *distance-2-simplicial* if the vertices of $N_2(v)$ form a distance-2-clique.

# 3 | RELATED WORK

In the following, we first discuss related work for the Maximum (Weight) 2-Packing Set problem in Section 3.1 and then, in Section 3.2, we cover recent work on the Maximum Weight Independent Set problem.

## 3.1 | 2-Packing Set Algorithms

In general, most of the contributions to the 2-packing set problem on arbitrary graphs are in the context of distributed algorithms for the maximal 2-packing set problem [23, 41, 46, 20, 50]. Ding *et al.* [17] propose a self-stabilizing algorithm on arbitrary graphs. The algorithm consists of two operations: entering and exiting the solution candidate for each vertex in the graph. If a vertex enters the solution, its neighbors get locked, so they can not enter the solution and cause a conflict. The decision to enter or exit the solution depends on whether a vertex causes a conflict. Mjelde [42] presents a self-stabilizing algorithm for the maximum $k$-packing set problem on tree graphs using dynamic programming.

For special graphs, there are further algorithms for the M2PS problem. Soto *et al.* [30] analyze the size of an M2PS for 2-token graphs of paths. Flores-Lamas *et al.* [19] present an algorithm that finds an M2PS in $O(n^2)$ time for an arbitrary cactus of order $n$. Trejo-Sánchez *et al.*[52] present an approximation algorithm for planar graphs using graph decompositions and LP-solvers. The approximation ratio is related to how the algorithm decomposes the input graph into smaller subgraphs, inspired by Baker [7].

Under the name of distance-3 independent or scattered set, there is further theoretical work for special graph classes [6, 35]. For arbitrary graphs, Yamanaka *et al.* [57] introduced a theoretical exact algorithm to solve the problem in $\mathcal{O}(1.4143^n)$ time. In their work, the authors use some simple reduction rules. However, these rules only reduce a branch-and-reduce instance consisting of a graph combined with a set of vertices that have to be in the solution and a set that is not.

There are only a few contributions to sequential, practical algorithms for the maximum 2-packing set problem on arbitrary graphs. Trejo-Sánchez *et al.*[49] are the first authors to have proposed a sequential algorithm for connected arbitrary graphs. They developed a genetic algorithm for the M2PS problem using local improvements in each round of their algorithm and a penalty function. Recently, Borowitz *et al.* [9] proposed the first exact data reduction rules for the M2PS problem. Additionally, the authors introduce an exact algorithm and a heuristic utilizing these data reduction rules to solve the M2PS problem for arbitrary graphs. These algorithms both outperformed the algorithm for general graphs by Trejo-Sánchez *et al.*[49] regarding solution quality and running time in multiple orders of magnitude.

Atsuta and Takahashi [4] introduce an approach considering the decision version of the Maximum Weight Distance-$d$ Independent Set problem in interval graphs. In contrast to the optimization problem that we are focusing on, in the decision version, the goal is to decide whether a distance-$d$ independent set of cardinality at least $k$ exists. To the best of our knowledge, there are no practical algorithms for the more general Maximum *Weight* 2-Packing Set problem. However, the possibility of solving the M2PS problem by using a graph transformation to the square graph and applying independent set solvers was first stated by Halldórsson *et al.*[32]. This approach is also applicable to the weighted case and is used in this work.

## 3.2 | Maximum Weight Independent Set Algorithms

Since one of our contributions is a transformation routine combined with different MWIS solvers, we summarize closely related work to these solvers here. However, we restrict this section to covering only the most recent state-of-the-art work that contributes to solving the MWIS problem in practice. The branch-and-reduce paradigm has effectively solved the M(W)IS and its complementary problem, the minimum vertex cover [3] to optimality. By this paradigm, we mean branching algorithms that use reduction rules to decrease the input size. Akiba and Iwata showed that this approach yields good results in comparison to other exact approaches for the minimum vertex cover and the maximum independent set problem [3]. Further, this approach is successfully used for the MWIS problem. Lamm *et al.*[37] use this approach for an exact algorithm KaMIS b&r, which is the current state-of-the-art for solving the MWIS problem exactly.

Further, data reductions are also useful for heuristics. One reduction-based heuristic, HtWIS, presented by Gu *et al.* [29], applies reductions exhaustively and then chooses one vertex by a tie-breaking policy to add to the solution. Moreover, Großmann *et al.* [28] use reduction rules in combination with an evolutionary approach for solving the MWIS problem on huge sparse networks.

Additionally, to solvers that use data reduction rules, we also want to mention a widely-known local search algorithm, the hybrid iterated local search, called HILS, by Nogueira *et al.* [43]. This heuristic uses efficient neighborhood swaps and random permutations to balance quality and diversity dynamically during the search. Recently, a new approach, the concurrent iterated local search, called CHILS, was introduced by Großmann *et al.* [26]. The authors utilize multiple local search solutions to compute a subgraph containing only the vertices whose solution status varies to focus the search. Local search is alternately applied to the original and the recomputed core instance.

# 4 | DATA REDUCTION RULES FOR THE MW2PS PROBLEM

After introducing basic concepts in Section 4.1, we present the 13 new data reduction rules for the MW2PS problem in Section 4.2. In Section 4.4, we present very efficient data reduction rules, which especially exploit the structure of a 2-packing set.

The reduction rules presented here are derived from data reduction rules for the Maximum Weight Independent Set problem presented in [29] and [37]. A comprehensive overview of all data reductions for the MWIS problem is given in [27].

The presented reduction rules are applied exhaustively in a predefined order in our preprocessing. Whenever a reduction rule is applied successfully, we check the previous rules again on the parts of the graph that changed. After applying a rule, we refer to the resulting instance as the *reduced* instance. After this step, when no reduction rules are applicable to the instance we pass it to the next step of the algorithm. We refer to this instance as $\mathcal{K}$.

All rules are introduced following the scheme as used in [27]. We first define the pattern in the graph that the corresponding rule can reduce. Then, we present the details of the construction of the reduced instance $\mathcal{G}'$. The reduced link-graph is constructed in two steps. First, we extend the *link set* $\mathcal{L}$ of the link-graph $\mathcal{G}$. In this step, for a given set of vertices $X \subset V$ that is removed from the link-graph in the corresponding reduction, we add links between two vertices $u, v \in V \setminus X$, if the shortest path between $u$ and $v$ is of length two in $\mathcal{G}$ however, it is longer in $\mathcal{G} - X$, i.e. all shortest paths between $u$ and $v$ included vertices in $X$. We have to add these links to maintain this necessary 2-neighborhood information linking the vertices $u$ and $v$ in the reduced link-graph.

Then, $\mathcal{G}'$ is defined by removing vertices from $\mathcal{G}$. The *offset* describes the difference between the weight of an MW2PS on the reduced instance $\alpha_w^2(\mathcal{G}')$ and the weight of an MW2PS on the original graph $\alpha_w^2(\mathcal{G})$. In the *reconstruction*, we present how the solution on the reduced instance $\mathcal{S}'$ is used to construct a solution $\mathcal{S}$ for the original graph.

## 4.1 | Basic Concepts

We first introduce a meta reduction Referencesred:meta, illustrated in Figure 1. It helps to understand the intuition behind the different reduction rules introduced later. Moreover, this example also shows the importance of our additional link set $\mathcal{L}$. Without the link added between the vertices $x$ and $y$ in the reduced instance, both of these vertices could be part of an MW2PS in the reduced instance. This would create a conflict, as $x$ and $y$ have a common neighbor in the original graph. The link set $\mathcal{L}$ ensures that the vertices $x$ and $y$ are not both part of an MW2PS in the reduced instance which is crucial for the correctness of our reduction rules.

**FIGURE 1** Illustration of Heavy Vertex. The left graph is the original link-graph $\mathcal{G}$. Dashed lines show links in $\mathcal{L}$. Red numbers indicate the vertex weights. On the reduced link-graph $\mathcal{G}'$, on the right, light edges, and vertices are reduced (green included and red excluded). Note that after the reduction $\deg(y) < \deg_L(y)$.

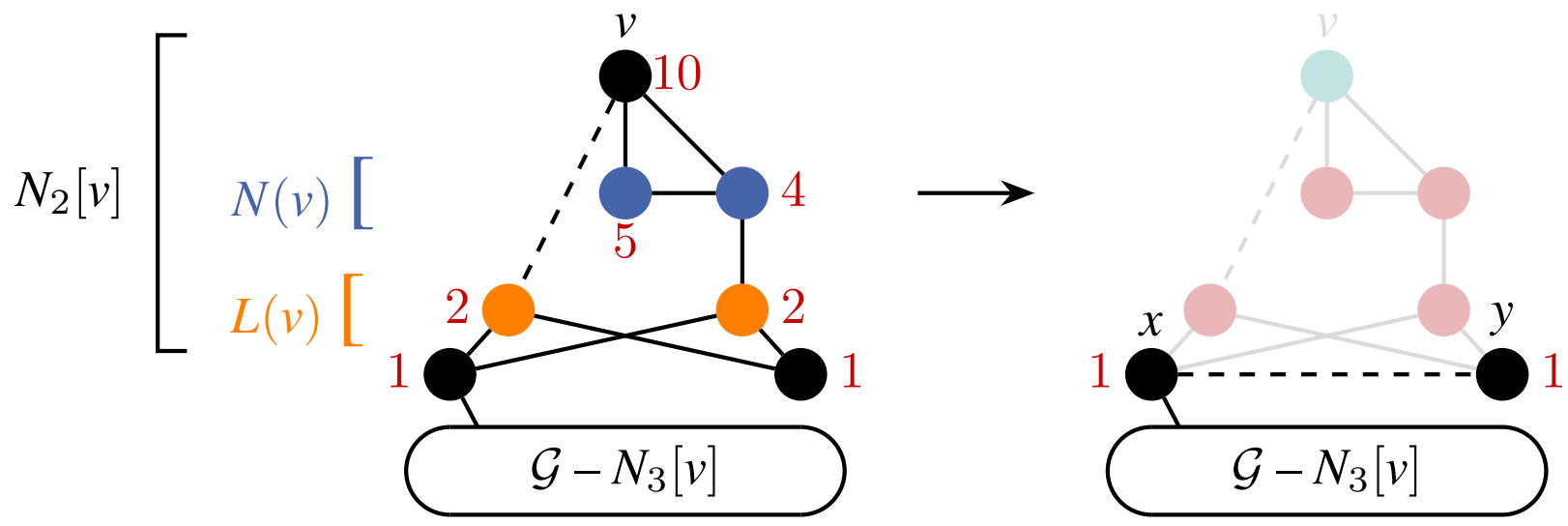

**Reduction 1** (Heavy Vertex)**.** *Let $v \in V$ such that the weight of the MW2PS in $N_2(v)$ is smaller than the weight of the vertex, i.e. $\alpha_w^2(\mathcal{G}[N_2(v)]) \leq w(v)$, then include $v$.*

| | |
|---|---|
| *Link Set* | $\mathcal{L} = \mathcal{L} \cup \{\{x, y\} \in \binom{N(N_2[v])}{2} \mid N(x) \cap N(y) \neq \emptyset\}$ |
| *Reduced Graph* | $\mathcal{G}' = \mathcal{G} - N_2[v]$ |
| *Offset* | $\alpha_w^2(\mathcal{G}) = \alpha_w^2(\mathcal{G}') + w(v)$ |
| *Reconstruction* | $\mathcal{S} = \mathcal{S}' \cup \{v\}$ |

*Proof.* Let $v \in V$ such that $\alpha_w^2(\mathcal{G}[N_2(v)]) \leq w(v)$ and $\mathcal{S}$ be the MW2PS. If $v \notin \mathcal{S}$, then we construct $\mathcal{S}' = \mathcal{S}' \setminus \{\mathcal{S} \cap N_2(v)\} \cup \{v\}$. The set $\mathcal{S}'$ is also a valid MW2PS and since $\alpha_w^2(\mathcal{G}[N_2(v)]) \leq w(v)$, it holds $w(\mathcal{S}) \leq w(\mathcal{S}')$. Therefore, we can always construct a solution including $v$. □

Considering the computational expense of calculating $\alpha_w^2(\mathcal{G}[N_2(v)])$, especially for large sets, we impose a bound on it instead. The most intuitive, also used in the independent set approaches, is summing up the weights in the set. It clearly holds that $\alpha_w^2(\mathcal{G}[N_2(v)]) \leq w(N_2(v))$. Given the nature of a 2-packing set, we can tighten this bound by the following observation. The direct neighborhood $N(v)$ forms a distance-2-clique since each neighbor has a common neighbor, namely $v$. Therefore, there can only be one of the direct neighbors contributing to $\alpha_w^2(\mathcal{G}[N_2(v)])$. We bound its weight by $w_{max}(N(v))$, yielding the following lemma.

**Lemma 1.** *Let $v \in V$ and $\deg(v) \geq 1$, then $N(v)$ forms a distance-2-clique and it holds*

$$\alpha_w^2(\mathcal{G}[N_2(v)]) \leq w(L(v)) + w_{max}(N(v)).$$

*Proof.* Let $v \in V$. The statement is clear if $\deg(v) = 1$. Let therefore $\deg(v) > 1$. Each neighbor in $N(v)$ is adjacent to $v$, and therefore, each link-neighbor shares a common neighbor, which yields $N(v)$ as a distance-2-clique. It is clear that $w(N_2(v)) = w(N(v)) + w(L(v)) \geq w_{max}(N(v)) + w(L(v))$. We now show that this is still a bound for $\alpha_w^2(\mathcal{G}[N_2(v)])$. The set $N(v)$ forms a distance-2-clique. Therefore, only one of its vertices can be part of a MW2PS. We can bound this by the maximum weight in the direct neighborhood. Therefore, it holds $w_{max}(N(v)) + w(L(v)) \geq \alpha_w^2(\mathcal{G}[N_2(v)])$. □

In the example illustrated in Figure 1, we have $\alpha_w^2(\mathcal{G}[N_2(v)]) = 7$ and Heavy Vertex is applicable. However, using the naive bound by summing up the weights $w(N_2(v)) = 13 \nleq 10 = w(v)$, we can not reduce it. Using Lemma 1 on the other hand, we get $w_{max}(N(v)) + w(L(v)) = 9 \leq 10 = w(v)$ and $v$ is reducible.

Figure 1 also highlights the importance of our link set. The orange vertices with common neighbors in the original graph are *linked* in the reduced link-graph. This example shows that in a reduced instance, there can be vertices $x, y \in V$ such that $\deg(y) < \deg_L(y)$. Without the additional link set, it is not possible to maintain the correct 2-neighborhood information of the vertices during the reduction process.

## 4.2 | Data Reduction Rules

Now we introduce the different data reduction rules used. The first reduction searches for neighbors $u \in N_2(v)$ of a vertex $v$, which can be removed since they can always be swapped for the vertex $v$ in a solution containing $u$. We can use different bounds depending on whether $u$ is a direct or a link neighbor.

**Reduction 2** (Neighbor Removal)**.** *Let $u, v \in V$ with*

1. *$u \in N(v)$ and $\alpha_w^2(\mathcal{G}[L(v) \setminus N_2[u]]) + w(u) \le w(v)$, or*
2. *$u \in L(v)$ and $\alpha_w^2(\mathcal{G}[N_2[v] \setminus N_2[u]]) + w(u) \le w(v)$,*

*then, exclude u.*

| | |
|---|---|
| *Link Set* | $\mathcal{L} = \mathcal{L} \cup \{\{x, y\} \in \binom{N(u)}{2}\}$ |
| *Reduced Graph* | $\mathcal{G}' = \mathcal{G} - u$ |
| *Offset* | $\alpha_w^2(\mathcal{G}) = \alpha_w^2(\mathcal{G}')$ |
| *Reconstruction* | $\mathcal{S} = \mathcal{S}'$ |

*Proof.* Let $v \in V$ and $u \in N_2[v]$ with $\alpha_w^2(\mathcal{G}[N_2[v] \setminus N_2[u]]) + w(u) \le w(v)$. In the case of $u \in N(v)$, it holds $N[v] \subseteq N_2[u]$ which results in $L(v) \setminus N_2[u] = N_2[v] \setminus N_2[u]$. Further, let $\mathcal{S}$ be an MW2PS of $\mathcal{G}$. If $u$ is not part of $\mathcal{S}$, then it is safe to remove $u$. Otherwise, it holds $u \in \mathcal{S}$. Then $v \in N_2[u]$ is not in $\mathcal{S}$ and $w(\mathcal{S} \cap N_2[v]) = w(\mathcal{S} \cap (N_2[v] \setminus N_2[u]) \cup \{u\}) = \alpha_w^2(\mathcal{G}[N_2[v] \setminus N_2[u]]) + w(u) = w(v)$; otherwise $u$ and $\mathcal{S} \cap N_2[v] \setminus N_2[u]$ can be swapped with $v$ in $\mathcal{S}$ for obtaining a 2-packing set of larger weight. Thus, $\mathcal{S}' = \mathcal{S} \setminus N_2[v] \cup \{v\}$ is a MW2PS of $\mathcal{G}$ excluding $u$ and $\alpha_w^2(G) = \alpha_w^2(G')$. □

The next reduction is illustrated in Figure 1 and combines the idea of Heavy Vertex and uses the inequality from Lemma 1 as a bound.

**Reduction 3** (Neighborhood Removal)**.** *Let $v \in V$. If $w(v) \ge w(L(v)) + w_{max}(N(v))$, then, include v.*

| | |
|---|---|
| *Link Set* | $\mathcal{L} = \mathcal{L} \cup \{\{x, y\} \in \binom{N(N_2[v])}{2} \mid N(x) \cap N(y) \ne \emptyset\}$ |
| *Reduced Graph* | $\mathcal{G}' = \mathcal{G} - N_2[v]$ |
| *Offset* | $\alpha_w^2(\mathcal{G}) = \alpha_w^2(\mathcal{G}') + w(v)$ |
| *Reconstruction* | $\mathcal{S} = \mathcal{S}' \cup \{v\}$ |

*Proof.* Let $v \in V$ and $w(v) \ge w(L(v)) + w_{max}(N(v))$. We want to apply Neighbor Removal. Therefore, we distinguish two cases. First, let $u \in N(v)$ and assume it is in some MW2PS. Then, $\alpha_w^2(\mathcal{G}[N_2[v] \setminus N_2[u]]) + w(u) = \alpha_w^2(\mathcal{G}[L(v) \setminus N_2[u]]) + w(u)$ and we can bound this by $\alpha_w^2(\mathcal{G}[L(v) \setminus N_2[u]]) + w(u) \le \alpha_w^2(\mathcal{G}[L(v)]) + w_{max}(N(v)) \le w(L(v)) + w_{max}(N(v)) \le w(v)$, using Lemma 1. This way, we can always swap $v$ for $u$ and remove all vertices in the direct neighborhood of $v$. Second, let $u \in L(v)$ be in some MW2PS. Since we already removed the complete direct neighborhood, it holds $\alpha_w^2(\mathcal{G}[N_2[v] \setminus N_2[u]]) + w(u) = \alpha_w^2(\mathcal{G}[L(v) \setminus N_2[u]]) + w(u) \le w(L(v) \setminus N_2[u]) + w(u)$. Similar to the first case, we show that $u$ can always be replaced by $v$ with the following estimation $w(L(v) \setminus N_2[u]) + w(u) \le w(L(v)) \le w(v)$. This argument allows us to reduce the complete 2-neighborhood, and $v$ is the only vertex left in its component. Hence, $v$ must be in an MW2PS and can be removed from $G$. □

In Split Neighbor Removal, we tighten the bound used in Neighbor Removal even further, depending on different special cases.

**Reduction 4** (Split Neighbor Removal)**.** *Let $u, v \in V$ such that $u \in N_2(v)$. If one of the following cases applies, we have an upper-bound U for $\alpha_w^2(\mathcal{G}[N_2[v] \setminus N_2[u]])$ and obtain a special case of the Neighbor Removal Reduction: $U + w(u) \le w(v)$.*

1. *If $u \in N(v)$, then, $U = w(L(v) \setminus N_2[u])$.*
2. *If $u \in L(v)$, then, $U = \min\{w(N_2[v] \setminus N_2[u]),\ w_{max}(N(v)) + w(L(v) \setminus N_2[u])\}$.*

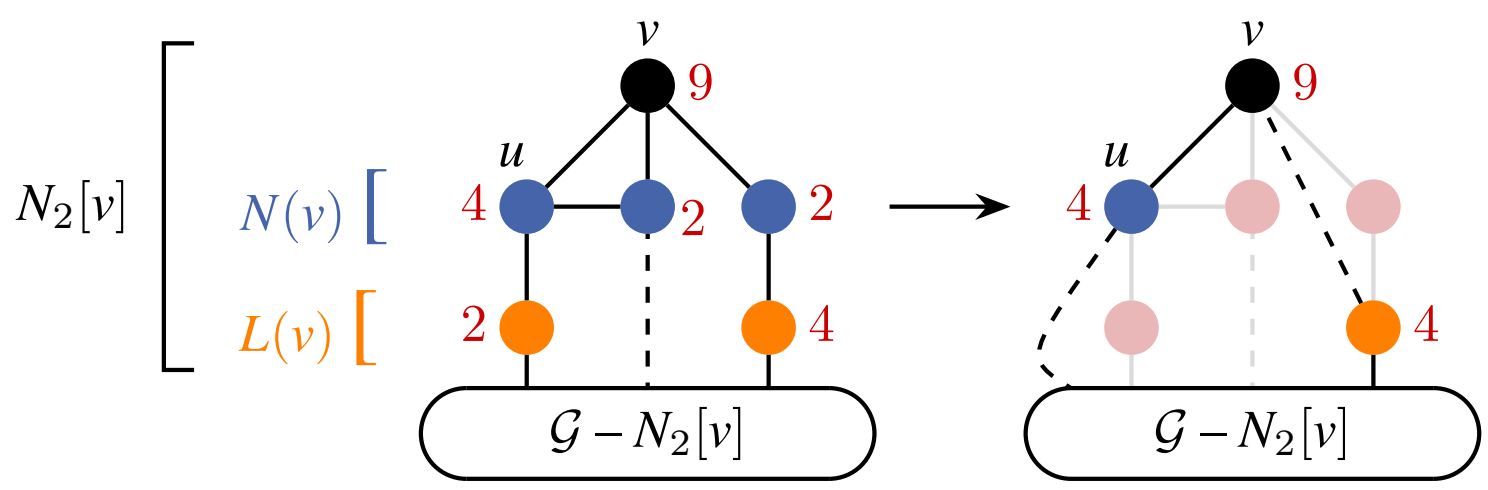

**FIGURE 2** Application of Split Intersection Removal. The original link-graph on the left is reduced to the link-graph on the right. Red numbers indicate vertex weights, dashed lines links in $\mathcal{L}$. Light-colored vertices and edges are reduced. Note that Intersection Removal is not applicable in this case.

*In these cases, we can exclude u.*

| | |
|---|---|
| *Link Set* | $\mathcal{L} = \mathcal{L} \cup \{\{x,y\} \in \binom{N(u)}{2}\}$ |
| *Reduced Graph* | $\mathcal{G}' = \mathcal{G} - u$ |
| *Offset* | $\alpha_w^2(\mathcal{G}) = \alpha_w^2(\mathcal{G}')$ |
| *Reconstruction* | $\mathcal{S} = \mathcal{S}'$ |

*Proof.* Let $u, v \in V$ such that $u \in N_2[v]$. We generally start with the upper-bound $U = w(\mathcal{G}[N_2[v] \setminus N_2[u]])$ from the Neighbor Removal Reduction. We can further tighten the bound and simplify its computation in the following cases.

1. Follows from Neighbor Removal (case 1).
2. Let $u \in L(v)$. Using Lemma 1, we estimate $\alpha_w^2(N_2[v] \setminus (N_2[u] \cup \{v\})) \leq w_{max}(N(v)) + w(L(v) \setminus N_2[u])$. Note that we can get a tighter bound when we stay with the original $w(\mathcal{G}[N_2[v] \setminus N_2[u]])$. We take the minimum of these two bounds.

□

*Remark 1. Efficient Split Neighbor Removal.* Before we compute $w(L(v) \setminus N_2[u])$ in Split Neighbor Removal, we check whether $w(N(u)) \geq w(L(v)) + w(N(v))$ and if true, exclude $u$. We can do this, since $w(N(u)) \geq w(L(v)) + w(N(v)) = w(N_2[v]) - w(v)$ is equivalent to $w(N_2[v]) - w(N(u)) \leq w(v)$. We get $w(L(v) \setminus N_2[u]) + w(u) \leq w(N_2[v] \setminus N(u)) = w(N_2[v]) - w(N(u)) \leq w(v)$ and can apply case 1 of Split Neighbor Removal. This yields a tighter bound $U = w(L(v) \setminus N_2[u])$ if $w(N(v) \setminus N[u]) < w(L(v) \cap N(u)) = w(N(u) \setminus N[v])$.

The idea behind the next reduction, called Intersection Removal, is similar to Neighbor Removal; however, we now consider two adjacent vertices: $u$ and $v$. If the reduction is applicable, one of these will be in the MW2PS; therefore, we can exclude their common neighbors.

**Reduction 5** (Intersection Removal)**.** *Let $u, v \in V$ such that $u \in N_2(v)$. If $w(v) \geq w(N_2(v) \setminus \{u\})$, then, exclude $K = (N_2[u] \cap N_2[v]) \setminus \{u, v\}$.*

| | |
|---|---|
| *Link Set* | $\mathcal{L} = \mathcal{L} \cup \{\{x,y\} \in \binom{N(K)}{2} \mid N(x) \cap N(y) \neq \emptyset\}$ |
| *Reduced Graph* | $\mathcal{G}' = \mathcal{G} - K$ |
| *Offset* | $\alpha_w^2(\mathcal{G}) = \alpha_w^2(\mathcal{G}')$ |
| *Reconstruction* | $\mathcal{S} = \mathcal{S}'$ |

*Proof.* Let $u, v \in V$ such that $u \in N_2(v)$ and $w(v) \geq w(N_2(v) \setminus \{u\})$. We assume a set of vertices $\mathcal{S} \subset N_2(v) \setminus \{u\}$ is part of an MW2PS. We know that $w(v) \geq w(N_2(v) \setminus \{u\}) \geq w(\mathcal{S})$. Therefore, we can always create a new solution $\mathcal{S}' = \mathcal{S} \setminus \mathcal{S} \cup \{v\}$ of larger or equal size. This shows that either $v$ or $u$ are in an MW2PS, so the intersection of their neighborhoods can be removed. □

Split Intersection Removal, illustrated in Figure 2, tightens the bounds used by the Intersection Removal by further using Lemma 1.

**Reduction 6** (Split Intersection Removal). *Let $u, v \in V$ and $K = (N_2(u) \cap N_2(v)) \setminus \{u, v\}$. If*

1. $u \in N(v)$, $w(v) \geq w(L(v)) + w_{max}(N(v) \setminus \{u\})$
2. $u \in L(v)$ *and* $w(v) \geq w(L(v) \setminus \{u\}) + w_{max}(N(v))$,

*then, exclude K.*

| | |
|---|---|
| *Link Set* | $\mathcal{L} = \mathcal{L} \cup \{\{x, y\} \in \binom{N(K)}{2} \mid N(x) \cap N(y) \neq \emptyset\}$ |
| *Reduced Graph* | $\mathcal{G}' = \mathcal{G} - K$ |
| *Offset* | $\alpha^2_w(\mathcal{G}) = \alpha^2_w(\mathcal{G}')$ |
| *Reconstruction* | $\mathcal{S} = \mathcal{S}'$ |

*Proof.* This reduction follows using Lemma 1 and Intersection Removal. □

The following reduction uses a special relation between two adjacent vertices $u$ and $v$. It is applicable if $N[u] = N_2[v]$. In this case, $N_2[v]$ forms a distance-2-clique, and $v$ can be included if it has the highest weight in $N_2[v]$, see Weighted Clique. If not, we exclude as many neighbors as possible.

**Reduction 7** (Domination). *Let $v \in V$ and $u \in N(v)$ with $N[u] = N_2[v]$. If*

1. $w(v) \geq w_{max}(N[u])$, *then, include v.*

| | |
|---|---|
| *Link Set* | $\mathcal{L} = \mathcal{L} \cup \{\{x, y\} \in \binom{N_2[v]}{2} \mid N(x) \cap N(y) \neq \emptyset\}$ |
| *Reduced Graph* | $\mathcal{G}' = \mathcal{G} - N[v]$ |
| *Offset* | $\alpha^2_w(\mathcal{G}) = \alpha^2_w(\mathcal{G}') + w(v)$ |
| *Reconstruction* | $\mathcal{S} = \mathcal{S}' \cup \{v\}$ |

2. $w(v) \geq w(N(u) \setminus \{v\})$, *then, exclude* $K = N(u) \setminus \{v\}$.

| | |
|---|---|
| *Link Set* | $\mathcal{L} = \mathcal{L} \cup \{\{x, y\} \in \binom{N(K)}{2} \mid N(x) \cap N(y) \neq \emptyset\}$ |
| *Reduced Graph* | $\mathcal{G}' = \mathcal{G} - K$ |
| *Offset* | $\alpha^2_w(\mathcal{G}) = \alpha^2_w(\mathcal{G}')$ |
| *Reconstruction* | $\mathcal{S} = \mathcal{S}'$ |

3. $w(v) \geq w(u)$, *then, exclude u.*

| | |
|---|---|
| *Link Set* | $\mathcal{L} = \mathcal{L} \cup \{\{x, y\} \in \binom{N(u)}{2}\}$ |
| *Reduced Graph* | $\mathcal{G}' = \mathcal{G} - u$ |
| *Offset* | $\alpha^2_w(\mathcal{G}) = \alpha^2_w(\mathcal{G}')$ |
| *Reconstruction* | $\mathcal{S} = \mathcal{S}'$ |

*Proof.* The first case holds since every neighbor of $v$ is a direct neighbor of $u$. Since $u$ spans a distance-2-clique over its direct neighborhood, $v$ is distance-2-simplicial. Therefore, if $w(v) \geq w_{max}(N[u])$, we can include in the solution. In the second case, if $w(v) \geq w(N(u) \setminus \{v\})$ we get $w(N(u) \setminus \{v\}) = w(N[u] \setminus \{u, v\}) \geq w(N_2[v] \setminus \{v, u\})$ and we can apply Intersection Removal to get the stated result. In the third case, if $w(v) \geq w(u)$, we can apply Split Neighbor Removal Case 1. □

*Remark 2. (Memory) Efficient Domination.* Instead of testing $N[u] = N_2[v]$, we test if $\deg_L(v) + \deg(v) = \deg(u)$, which is equivalent, since for $u \in N(v)$, each neighbor of $u$ is a link-neighbor of $v$, i.e. $N[u] \subseteq N_2[v]$. The degree equality yields that $|N[u]| = 1 + \deg(u) = 1 + \deg_L(v) + \deg(v) = |N_2[v]|$. Therefore, the sets have to be equal, and we get $N[u] = N_2[v]$. To check this degree equality, we need to initialize the link-neighborhood of each vertex. To avoid this for some cases when Domination is not applicable and the link-degree not computed yet, we first check whether $N[v] \subseteq N[u]$ and only compute the link-neighborhoods for these vertices.

The following reduction rule, Weighted Clique includes vertices $v$ if these are distance-2-simplicial,i.e. their neighborhood $N_2[v]$ forms a distance-2-clique. If the weight of $v$ is the highest in $N_2[v]$, we can include $v$ in the solution.

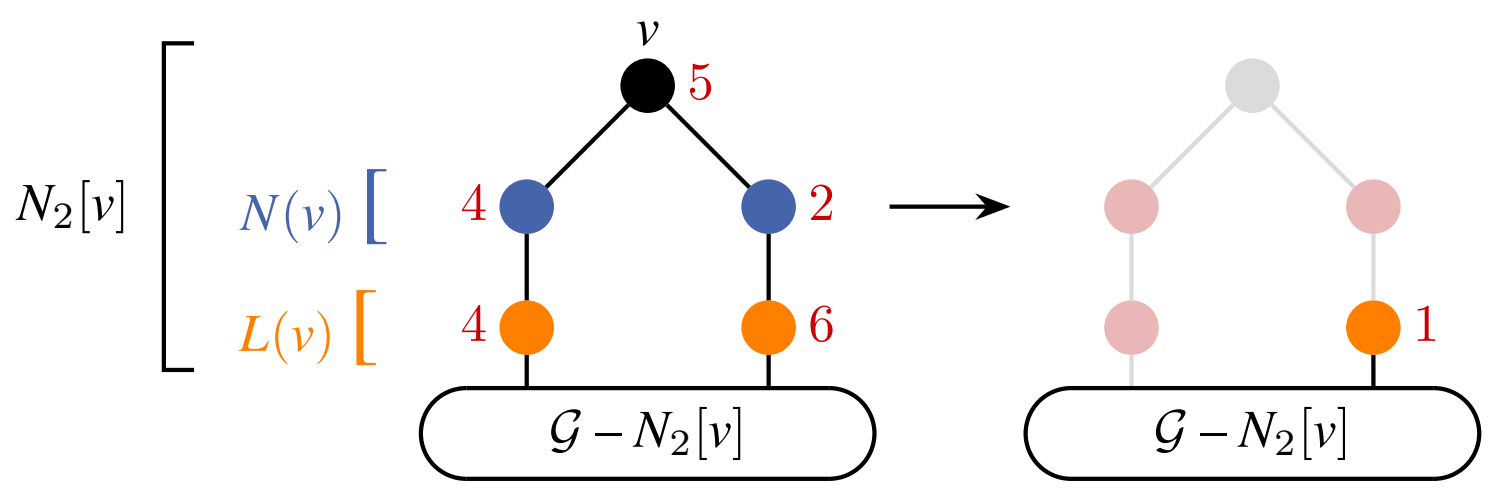

**FIGURE 3** Application of D2-Simplicial Weight Transfer. The original link-graph on the left is reduced to the link-graph on the right. Red numbers indicate vertex weights. Light red vertices are excluded, gray vertices are folded, and light gray edges are removed.

**Reduction 8** (Weighted Clique)**.** *Let $v \in V$ be distance-2-simplicial in $\mathcal{G}$ such that $w(v) = w_{max}(N_2[v])$, then, include $v$.*

| | |
|---|---|
| *Link Set* | $\mathcal{L} = \mathcal{L} \cup \{\{x,y\} \in \binom{N(N_2[v])}{2} \mid N(x) \cap N(y) \neq \emptyset\}$ |
| *Reduced Graph* | $\mathcal{G}' = \mathcal{G} - N_2[v]$ |
| *Offset* | $\alpha_w^2(\mathcal{G}) = \alpha_w^2(\mathcal{G}') + w(v)$ |
| *Reconstruction* | $\mathcal{S} = \mathcal{S}' \cup \{v\}$ |

*Proof.* Let $v \in V$ be distance-2-simplicial in $\mathcal{G}$ such that $w(v) = w_{max}(N_2[v])$. Since the vertex $v$ is distance-2-simplicial, i.e. the neighborhood $N_2[v]$ forms a distance-2-clique, there can only be one of these vertices in the MW2PS. Since for all vertices $u \in N_2[v]$ it holds that $\alpha_w^2(\mathcal{G}[N_2[v] \cup N_2[u]]) \leq \alpha_w^2(\mathcal{G}[N_2[v]]) = w_{max}(N_2[v]) = w(v)$, we can remove $u$ by applying the Neighborhood Removal Reduction resulting in $v$ being left to include in the solution. □

If Weighted Clique is not applicable, we can still reduce some part of the neighborhood of distance-2-simplicial vertices. The cases where this is possible are described in D2-Simplicial Weight Transfer, illustrated in Figure 3.

**Reduction 9** (D2-Simplicial Weight Transfer)**.** *Let $v \in V$ be distance-2-simplicial, and suppose that the set of distance-2-simplicial vertices $S^2(v) \subset N_2(v)$ is such that for all $u \in S^2(v)$ holds $w(v) \geq w(u)$, then fold $v$ into its neighborhood.*

| | |
|---|---|
| *Link Set* | $\mathcal{L} = \mathcal{L} \cup \{\{x,y\} \in \binom{N(K)}{2} \mid N(x) \cap N(y) \neq \emptyset\}$ |
| *Reduced Graph* | *Construct the link-graph $\mathcal{G}'$ as follows* |

- *remove all $u \in K := \{u \in N_2(v) \mid w(u) \leq w(v)\}$, and let the remaining neighbors be denoted by $N'(v)$*
- *remove $v$ and for each $u \in N'(v)$ set $w(u) = w(u) - w(v)$*

| | |
|---|---|
| *Offset* | $\alpha_w^2(\mathcal{G}) = \alpha_w^2(\mathcal{G}') + w(v)$ |
| *Reconstruction* | *If $\mathcal{S}' \cup N'(v) = \emptyset$, then $\mathcal{S} = \mathcal{S}' \cup \{v\}$, else $\mathcal{S} = \mathcal{S}'$.* |

*Proof.* The first step of the D2-Simplicial Weight Transfer Reduction is applying the Neighbor Removal Reduction to all vertices $u \in N_2(v)$ with $w(u) \leq w(v)$. Let $x \in N'(v)$ and $\mathcal{S}'$ be the solution on $\mathcal{G}'$. In the second step, we need to distinguish two cases.

*Case 1:* $\mathcal{S}' \cap N'(v) = \emptyset$. We show that $w(v) + \alpha_w^2(\mathcal{G}[V \setminus N_2[v]]) \geq \alpha_w^2(\mathcal{G}[V \setminus \{v\}])$. Since $x \notin \mathcal{S}$, we have $\alpha_w^2(\mathcal{G}') \geq w'(x) + \alpha_w^2(\mathcal{G}'[V' \setminus N_2[x]]) = w(x) - w(v) + \alpha_w^2(\mathcal{G}'[V' \setminus N_2[x]])$. Therefore, the heaviest 2PS containing $v$ has at least the weight of the heaviest 2PS containing any link-neighbor of $v$. Since additionally, $w(v) + \alpha_w^2(\mathcal{G}[V \setminus N_2[v]]) \geq \alpha_w^2(\mathcal{G}[V \setminus \{v\}])$, we know that $\mathcal{S} = \mathcal{S}' \cup \{v\}$ is a MW2PS of $\mathcal{G}$.

*Case 2:* $\mathcal{S}' \cap N'(v) \neq \emptyset$. We show that $\mathcal{S}' = \mathcal{S}$ is a MW2PS of $\mathcal{G}$. Since $v$ is distance-2-simplicial, it holds $|\mathcal{S}' \cap N'(v)| = 1$. Define $\mathcal{G}''$ as the link-graph resulting from increasing the weights of $N'(v)$ again by $w(v)$, i.e. the original weights. The set $\mathcal{S}'$ is also an MW2PS for the link-graph $\mathcal{G}''$, since otherwise, it would not have been optimal on $\mathcal{G}'$. Therefore, we have $w'(\mathcal{S}') + w(v) = w(\mathcal{S}')$, since exactly one vertex from $N'(v)$ is part of $\mathcal{S}'$. In the next step, we add the vertex $v$ to the link-graph $\mathcal{G}''$, resulting in the link-graph $\mathcal{G}'''$, i.e. this is the link-graph after step 1 of the reduction. We now assume, that there is a MW2PS $\mathcal{S}^*$ for $\mathcal{G}'''$ with $w(\mathcal{S}^*) > w(\mathcal{S}')$. Then, $v \in \mathcal{S}^*$, since this is the only vertex added to $\mathcal{G}'''$. This implies that no neighbor of $v$ is in $\mathcal{S}^*$. We now have $w(\mathcal{S}^* \setminus \{v\}) = w(\mathcal{S}^*) - w(v) > w(\mathcal{S}') - w(v) = w'(\mathcal{S}') + w(v) - w(v) = w'(\mathcal{S}')$. Since $\mathcal{S}^* \setminus \{v\}$ does not contain any vertex from $N_2[v]$, it is an MW2PS of $\mathcal{G}'$ that is of larger weight than $\mathcal{S}'$. This contradicts the assumption, and we have $\mathcal{S}' = \mathcal{S}$. □

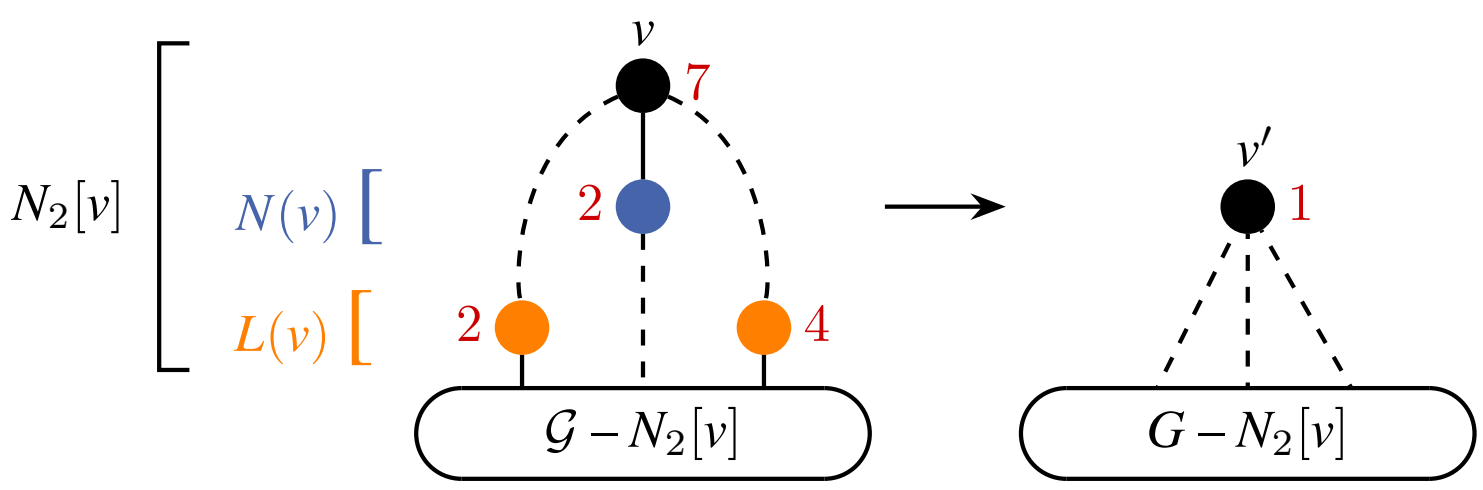

**FIGURE 4** Application of Neighborhood Folding. The original link-graph on the left is reduced to the instance on the right. Red numbers indicate vertex weights and dashed lines links in $\mathcal{L}$.

The following reduction describes a folding. Here, we postpone the decision about the solution status of a vertex to a later point. Depending on the solution status of the new vertex, we decide the solution status of the folded vertices.

Our Neighborhood Folding is applicable to a vertex $v$ with a 2-packing neighborhood, i.e. $N_2(v)$ is a 2-packing set in $\mathcal{G}$. In this situation, we cannot include the vertex $v$ directly, but we know that in the end, either $v$ or its complete neighborhood is in the solution. Therefore, we can fold these vertices. Neighborhood Folding is illustrated in Figure 4.

**Reduction 10** (Neighborhood Folding). *Let $v \in V$ and $N_2(v)$ be a 2-packing set in $\mathcal{G}$. If $w(N_2(v)) > w(v)$, but $w(N_2(v)) - w_{min}(N_2(v)) \leq w(v)$, then, fold $N_2[v]$ into a new vertex $v'$.*

| | |
|---|---|
| *Link Set* | $\mathcal{L} = \mathcal{L} \cup \{\{v', x\} \mid x \in N(N_2[v])\}$ |
| *Reduced Graph* | $\mathcal{G}' = \mathcal{G}[(V \setminus N_2[v]) \cup \{v'\}]$, *set weight* $w(v') = w(N_2(v)) - w(v)$ *and* $N(v') = \emptyset$. |
| *Offset* | $\alpha_w^2(\mathcal{G}) = \alpha_w^2(\mathcal{G}') + w(v)$ |
| *Reconstruction* | *If* $v' \in \mathcal{S}'$ *then* $\mathcal{S} = (\mathcal{S}' \setminus \{v'\}) \cup (N_2(v))$, *else,* $\mathcal{S} = \mathcal{S}' \cup \{v\}$. |

*Proof.* Let $v \in V$ and $N_2(v)$ be a 2-packing set of $\mathcal{G}$. First, note that after folding, the following graphs are identical: $\mathcal{G}'[V' \setminus N_{2\mathcal{G}'}[v']] = \mathcal{G}[V \setminus N_4[v]]$ and $\mathcal{G}'[V' \setminus \{v'\}] = \mathcal{G}[V \setminus N_2[v]]$. Let $\mathcal{S}'$ be a MW2PS of $\mathcal{G}'$. Now, we distinguish two cases.

*Case 1: $v' \in \mathcal{S}'$.* We show that $w(N_2(v)) + \alpha_w^2(\mathcal{G}[V \setminus N_4[v]]) \geq w(v) + \alpha_w^2(\mathcal{G}[V \setminus N_2[v]])$. This shows that all neighbors of $v$ are together in a MW2PS of $\mathcal{G}$. Since $v' \in \mathcal{S}'$,

$$\begin{aligned}\alpha_w^2(\mathcal{G}') &= w(v') + \alpha_w^2(\mathcal{G}'[V' \setminus N_{2\mathcal{G}'}[v']]) \\ &= w(N_2(v)) - w(v) + \alpha_w^2(\mathcal{G}'[V' \setminus N_{2\mathcal{G}'}[v']]) \\ &= w(N_2(v)) - w(v) + \alpha_w^2(\mathcal{G}[V \setminus N_4[v]])\end{aligned}$$

Since $v' \in \mathcal{S}'$, we have $\alpha_w^2(\mathcal{G}') \geq \alpha_w^2(\mathcal{G}'[V' \setminus \{v'\}]) = \alpha_w^2(\mathcal{G}[V \setminus N_2[v]])$. Thus, $w(N_2(v)) + \alpha_w^2(\mathcal{G}[V \setminus N_4[v]]) \geq w(v) + \alpha_w^2(\mathcal{G}[V \setminus N_2[v]])$ and the vertices of the link-neighborhood are together in some MW2PS of $\mathcal{G}$ and $\alpha_w^2(\mathcal{G}) = w(N_2(v)) + \alpha_w^2(\mathcal{G}[V \setminus N_4[v]]) = \alpha_w^2(\mathcal{G}') + w(v)$.

*Case 2: $v' \notin \mathcal{S}'$.* We show that $v$ is in some MW2PS, by proving that $w(v) + \alpha_w^2(\mathcal{G}[V \setminus L(v)]) \geq w(N_2(v)) + \alpha_w^2(\mathcal{G}[V \setminus N_4[v]])$. Since $v' \notin \mathcal{S}'$, we have $\alpha_w^2(\mathcal{G}') = \alpha_w^2(\mathcal{G}'[V' \setminus \{v'\}]) = \alpha_w^2(\mathcal{G}[V \setminus N_2[v]])$. Since $\mathcal{S}'$ is a MW2PS of $\mathcal{G}'$, we obtain $\alpha_w^2(\mathcal{G}[V \setminus N_2[v]]) = \alpha_w^2(\mathcal{G}') \geq w(N_2(v)) - w(v) + \alpha_w^2(\mathcal{G}[V \setminus N_4[v]])$. Therefore, we have $w(v) + \alpha_w^2(\mathcal{G}[V \setminus L(v)]) \geq w(N_2(v)) + \alpha_w^2(\mathcal{G}[V \setminus N_4[v]])$ and $v$ is in some MW2PS of $\mathcal{G}$. Lastly, $\alpha_w^2(\mathcal{G}) = w(v) + \alpha_w^2(\mathcal{G}[V \setminus N_2[v]]) = w(v) + \alpha_w^2(\mathcal{G}')$. □

*Remark 3.* In Neighborhood Folding, we can efficiently prune the search space by checking if $\deg(v) \leq 1$. Otherwise, $N_2(v)$ would contain a distance-2-clique, see Lemma 1 and therefore not be a 2-packing set.

## 4.3 | Data Structure

In this section, we describe the data structure and some implementation details for the data reduction rules.

*Maintain Neighborhood Information.* We maintain important neighborhood information of a vertex to speed up the reduction process. For this, we maintain the size of the link-neighborhood $\deg_{\mathrm{L}}(v)$ and the weight of the maximum weight vertex. Additionally, we maintain the sum of the weights in the direct neighborhood for a vertex.

*On-Demand-Neighborhood.* Our reductions operate on a dynamic link-graph data structure based on adjacency arrays representing undirected edges and links by two directed ones. Internally, we separate edges and links with two adjacency arrays. In addition to the necessary links for the reductions, we compute and store the full link-neighborhood of a vertex whenever we compute it. We do this even if the reduction was not successful to avoid recomputation. Generally, we only compute the link-neighborhood of a vertex on demand. Here, we make special use of reductions that do not need to know the link-neighborhood to exclude a vertex. These reductions are applied initially, reducing the link-graph without building the full square graph. This approach additionally reduces the amount of work needed when reducing the instance. When a vertex $u$ is removed from the link-graph, we delete every edge and all links pointing to it. Removing all incoming edges can take time $\mathcal{O}(\Delta^4)$ where $\Delta$ is the highest degree in the graph. With the on-demand technique, the link-neighborhood of a vertex $v$ is not computed in advance, and therefore, potentially fewer edges must be deleted.

*Bulk Hide Operation.* Hiding the incoming edges and links can be computationally expensive for high-degree vertices. For example, for large distance-2-cliques, it can be expensive to exclude multiple clique members, i.e. applying Weighted Clique or D2-Simplicial Weight Transfer. Hiding one member of this distance-2-clique takes $\mathcal{O}(\Delta^4)$ time because we have to search through the edge and link adjacency array for each clique member to remove the connections. Then, hiding $k \in \mathcal{O}(\Delta^2)$ distance-2-clique members this takes $\mathcal{O}(k\Delta^4)$ time. To avoid this, we put all the hide operations of the respective distance-2-clique members together by using a *bulk hide operation*. This operation hides all edges and links to clique members in a single pass. This reduces the running time from $\mathcal{O}(k\Delta^4)$ to $\mathcal{O}(\Delta^4)$ and possibly decreases cache faults.

## 4.4 | Efficient Data Reduction Rules

So far, we presented data reduction rules based on the link-neighborhood information. However, using the link-neighborhood, e.g. determining or iterating over the 2-neighborhood, can be expensive regarding running time and memory consumption. In our implementation of the reduction rules, we postpone determining a link neighborhood as long as possible. Especially if a reduction is not applicable, determining the link-neighborhood early results in more work throughout the whole reduction process as this link-neighborhood has to be maintained.

In this section, we present efficient data reduction rules which circumvent these issues. Whereas arbitrary link neighborhoods are of size $\mathcal{O}(\Delta^2)$, the following data reduction rules fully circumvent considering link neighborhoods of size $\omega(\Delta)$ or ensure that applying a data reduction rule for all vertices does overall take at most $\mathcal{O}(n+m)$ running time. Moreover, we implement them such that no link neighborhoods are initialized and maintained by the link-graph data structure during the applicability tests.

Instead, the main idea behind these rules is to exploit the properties of the vertices $V_G$ in the original input graph $G$ to reduce $\mathcal{G}$ further. Intuitively, we reduce vertices in the link-graph $\mathcal{G}$ by looking up neighborhoods of (already reduced) vertices in $G$. Note that $G$ refers to the original, unreduced input graph (without the link set $\mathcal{L}$). With $V_G$, we refer to vertices in the original graph $G$, while $N_G(v)$ denotes the original neighborhood of the vertex $v$ in $G$. We refer to neighborhoods in the currently reduced link-graph $\mathcal{G}$ if a refernce is omitted in the notation.

The first data reduction rule, Fast Degree-1, fully reduces vertices of degree one in the input graph $G$. When applying this reduction rule and building the link neighborhood in $\mathcal{G}$, a degree one vertex always forms a distance-2-clique. Figure 5 gives an example and points out the difference to the more general D2-Simplicial Weight Transfer Reduction, which also reduces this pattern. Fast Degree-1 can be applied to degree one and zero vertices that arise during the reduction progress in $\mathcal{G}$. Figure 5 illustrates Fast Degree-1.

**Reduction 11** (Fast Degree-1)**.** *Let $u \in V_G$ with $v \in V \cap N_G(u)$ so that $\deg(v) \leq 1$ and $L(v) \subset N_G(u)$. Further, let $T(v) = \{x \in V \cap N_G(u) : \deg(x) \leq 1 \land L(x) \subset N_G(u)\} \setminus \{v\}$ be denoted as the twins of $v$ in $\mathcal{G}$ with $w(v) \geq \max_{x \in T(v)} w(x)$ and at most degree 1. Then, fold $v$ and its twins.*

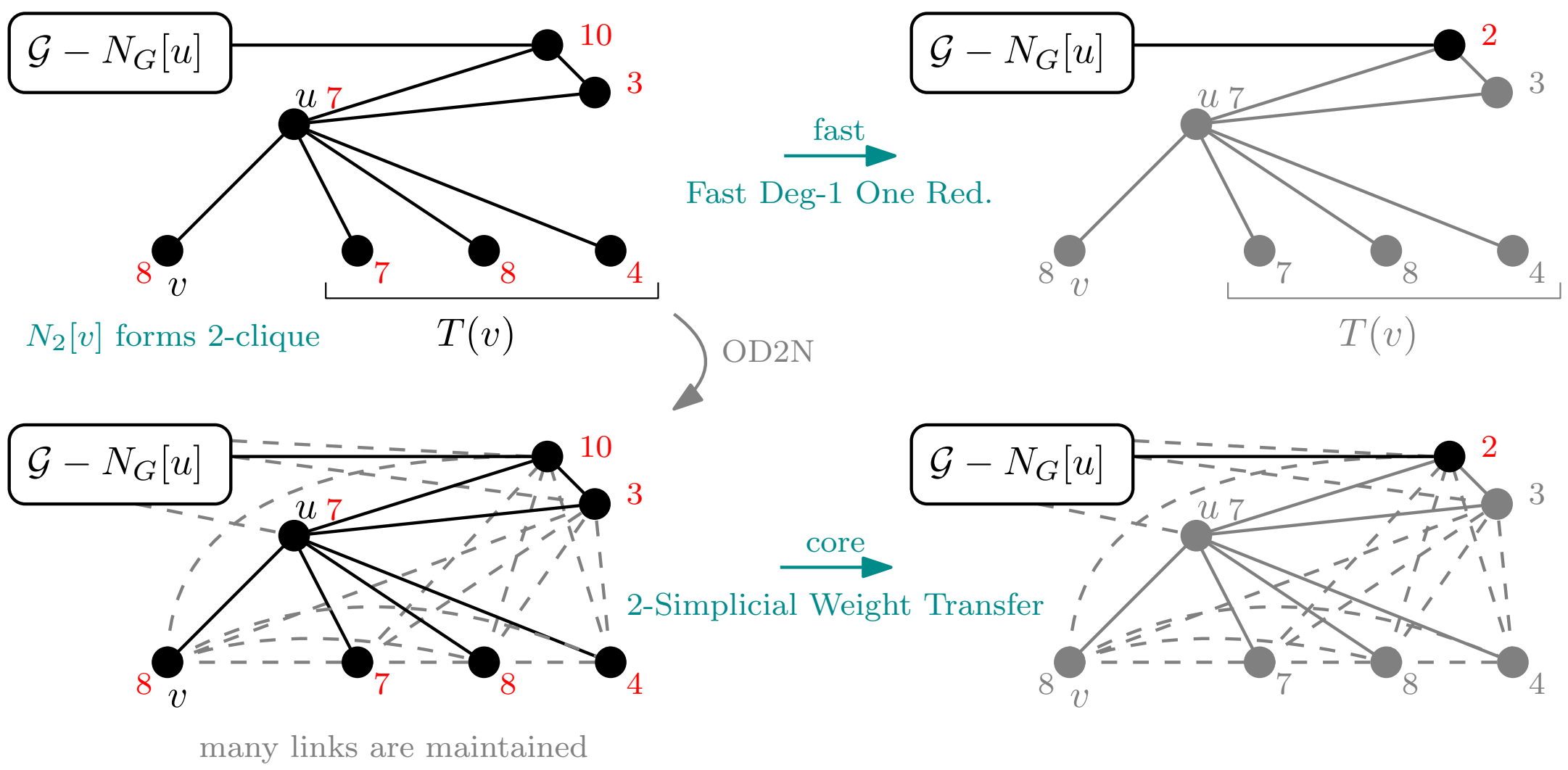

**FIGURE 5** This reduction example illustrates the benefits of using these fast reductions. Here, we display all the links that have to be computed during the reduction application, not only the ones necessary for the correctness of the reduced link-graph in the end. The Fast Degree-1 Reduction and the D2-Simplicial Weight Transfer are equally effective here, but in contrast to D2-Simplicial Weight Transfer, Fast Degree-1 fully circumvents to maintain links. Without Fast Degree-1, the links are initialized (OD2N) at the latest during D2-Simplicial Weight Transfer to check if $N_2[v]$ forms a distance-2-clique. When applying D2-Simplicial Weight Transfer, we have to hide many links, which is computationally expensive even when using the bulk hide operation. The Fast Degree-1, on the other hand, ensures that for $v$, the condition $L(v) \subset N(u)$ is true, and we know that $\deg(v) \leq 1$. Therefore, $v$ forms a distance-2-clique.

| | |
|---|---|
| *Link Set* | $\mathcal{L} = \mathcal{L}$ |
| *Reduced Graph* | *Construct the link-graph $\mathcal{G}'$ as follows* |

- *remove all $x \in (V \cap N_G[u]) \setminus \{v\}$ with $w(x) \leq w(v)$, and let the remaining link neighborhood of v be denoted by $L'(v)$*
- *remove v and for each $x \in L'(v)$ set $w(x) = w(x) - w(v)$*

| | |
|---|---|
| *Offset* | $\alpha_w^2(\mathcal{G}) = \alpha_w^2(\mathcal{G}') + w(v)$ |
| *Reconstruction* | *If $\mathcal{S}' \cup L'(v) = \emptyset$, then $\mathcal{S} = \mathcal{S}' \cup \{v\}$, else $\mathcal{S} = \mathcal{S}'$.* |

*Proof.* $T(v)$ is a subset of distance-2-simplicial vertices $S^2(v)$ of the considered distance-2-clique. We can now apply D2-Simplicial Weight Transfer by observing that we can relax its maximality constraint for $w(v)$. Although $v$ might not have maximal weight among all distance-2-simplicial vertices in this distance-2-clique, we can still apply a weight shift by $w(v)$. To see this, we assume that simplicial vertices of larger weight than $w(v)$ remain in $\mathcal{G}'$. One can apply D2-Simplicial Weight Transfer to the remaining distance-2-clique $L'(v)$ with a simplicial vertex $x \in L'(v)$ that satisfies the maximality constraint in $(\mathcal{G}', w')$, and check that this gives the same reduced instance and weight function as applying the D2-Simplicial Weight Transfer directly to $N_2[x]$ in $(\mathcal{G}, w)$. Note that $x$ also satisfies the maximality constraint in $\mathcal{G}$ since $w(x) = w'(x) + w(v) \geq w'(y) + w(v) = w(y)$ for all distance-2-simplicial $y \in S(v) \setminus T(v)$, and $w(x) \geq w(v) \geq w(y)$ for $y \in T(v)$. □

*Remark 4. Implementing the Fast Degree-1.* The implementation works in rounds where each round considers only degree one vertices. Initially, we consider all degree one vertices of $G$, as they remain distance-2-simplicial as long as they are not yet reduced. In the upcoming rounds, we reduce new degree one and zero vertices. Note that the subset condition for the link set is not trivially fulfilled for these vertices. Therefore, we maintain a list of vertices that do not fulfill that condition.

The following data reduction reduces degree-two vertices in $V_G$ almost analogously to Fast Degree-1. The key difference is that they do not necessarily point to distance-2-cliques. However, we can still reduce twins and possibly include a degree-2 vertex using Lemma 1.

**Reduction 12** (Fast Degree-2)**.** *Let $u, y \in V_G$ with $v \in V \cap N_G(u) \cap N_G(y)$ so that $\deg(v) \leq 2$ and $L(v) \subset N_G(u) \cup N_G(y)$. Further, let $T(v) = \{x \in V \cap N_G(u) \cap N_G(y) : \deg(x) \leq 2 \wedge L(x) \subset N_G(u) \cup N_G(y)\} \setminus \{v\}$ be denoted as the twins of v in $\mathcal{G}$ of at most degree two. If $w(v) \geq \max_{x \in T(v)} w(x)$, then fold v with its twins.*

| | |
|---|---|
| *Link Set* | *If v is included, $\mathcal{L} = \mathcal{L} \cup \{\{x, y\} \in \binom{N(N_2[v])}{2} \mid N(x) \cap N(y) \neq \emptyset\}$, else $\mathcal{L} = \mathcal{L}$* |
| *Reduced Graph* | *Construct the link-graph $\mathcal{G}'$ as follows* |

- *remove all $x \in T(v)$*
- *include v if $u \in N_G(y)$ and $w(v) \geq \max\{w(u), w(y), c_u + c_y\}$, or $u \notin N_G(y)$ and $w(v) \geq \max\{w(u) + c_y, w(y) + c_u, c_u + c_y\}$ where $c_y = \max_{x \in N(y) \setminus \{u,v\}} w(x)$ and $c_u = \max_{x \in N(u) \setminus \{y,v\}} w(x)$.*

| | |
|---|---|
| *Offset* | *$\alpha_w^2(\mathcal{G}) = \alpha_w^2(\mathcal{G}') + w(v)$ if v was included; otherwise $\alpha_w^2(\mathcal{G}) = \alpha_w^2(\mathcal{G}')$* |
| *Reconstruction* | *If $\mathcal{S} \cup L'(v) = \emptyset$, then $\mathcal{S} = \mathcal{S}' \cup \{v\}$, else $\mathcal{S} = \mathcal{S}'$.* |

*Proof.* Consider $u, y \in V_G$ with $v \in V \cap N_G(u) \cap N_G(y)$ as given above. Then, it holds $N(v) = \{u, y\} \cap V = N(x)$ for every $x \in T(v)$. Since all links incident to $v$ or $x$ are given via their (former) direct neigbors $u$ and $y$, it holds $L(x) \setminus \{v\} = L(v) \setminus \{x\}$. Therefore, we know $N_2[v] = N_2[u]$. If now $w(v) \geq \max_{x \in T(v)} w(x)$, then it holds $w(v) \geq w(x) = w(x) + w(N_2[v] \setminus N_2[x])$ for every twin $x \in T(v)$. Thus, we can apply the second case of Split Neighbor Removal to remove all twins safely.

If $v$ has sufficiently large weight, we can include $v$ by utalizing an upper bound $U$ for $\alpha_w^2(\mathcal{G}[N_2(v)])$. It allows us to apply Heavy Vertex and further to include $v$. The key observation is that the remaining vertices of $N_G(u) \cap V \setminus \{y\}$ form a distance-2-clique in $\mathcal{G}$. Analogously, $N_G[y] \cap V \setminus \{u\}$ forms a distance-2-clique in $\mathcal{G}$. Note that, at most one vertex of a distance-2-clique can participate in an MW2PS.

Now, consider that $u$ and $y$ are neighbors in $G$ and $w(v) \geq \max\{w(u), w(y), c_u + c_y\}$. Now, if $u$ and $y$ are direct neighbors, either $u$, $y$, or neighbors of $u$ or $y$ can be part of an MW2PS. Thus, $\max\{w(u), w(y), c_u + c_y\}$ is an upper bound for the 2-neighborhood of $v$.

If $u$ and $y$ are not adjacent, they are linked, and we can include $v$ if $w(v) \geq \max\{w(u) + c_y, w(y) + c_u, c_u + c_y\}$ with a similar argument. If $u$ is part of an MW2PS, only vertices of $N_G(y) \cap V$ regarding the 2-neighborhood of $v$ can be part of an MW2PS. Analogously, if $y$ is part of an MW2PS, only vertices of $N_G(u) \cap L(v)$ but none of $L(v) \setminus N_G(u)$ can be in an MW2PS. This gives us the upper bound for $\alpha_w^2(\mathcal{G}[N_2(v)])$, which allows us to include $v$ with Heavy Vertex.

□

*Remark 5. Implementing Fast Degree-2.* Fast Degree-2 can be implemented similarly to Fast Degree-1. However, we do not trace new vertices of degree two or one and test this reduction only once for each vertex of $V_G$.

The following data reduction is based on Neighborhood Removal and uses an upper bound on the summed weight in the link neighborhood. The upper bound is computed for $G$. Before any reductions are applied, we compute the upper bound for all vertices in time $\mathcal{O}(|E_G|)$ with two scans over all neighborhoods in $G$.

**Reduction 13** (Fast Neighborhood Removal)**.** *Let $v \in V$ with $w(v) \geq w_{max}(v) + \sum_{u \in N_G(u)} w(N_G(u)) - w(v)$, then include v.*

| | |
|---|---|
| *Link Set* | $\mathcal{L} = \mathcal{L} \cup \{\{x, y\} \in \binom{N(N_2[v])}{2} \mid N(x) \cap N(y) \neq \emptyset\}$ |
| *Reduced Graph* | $\mathcal{G}' = \mathcal{G} - N_2[v]$ |
| *Offset* | $\alpha_w^2(\mathcal{G}) = \alpha_w^2(\mathcal{G}') + w(v)$ |
| *Reconstruction* | $\mathcal{S} = \mathcal{S}' \cup \{v\}$ |

*Proof.* Let $v \in V$ with $w(v) \geq w_{max}(v) + \sum_{u \in N_G(u)} w(N_G(u)) - w(v)$. It holds $L(v) \subseteq N_G(N_G(v)) = \bigcup_{u \in N_G(v)} N_G(u) \setminus \{v\}$. Thus, we obtain $w(L(v)) \leq \sum_{u \in N_G(v)} w(N_G(u) - w(v))$ and can apply Neighborhood Removal. □

# 5 | SOLVING THE MW2PS PROBLEM

In this section, we focus on solving the MW2PS problem using the data reduction rules introduced in the previous section. First, we present the reduce and transform routine, which enables us to use well-studied independent set solvers to solve the MW2PS problem. Second, we introduce our new heuristic algorithm redW2pack which applies exact and heuristic data reductions to solve the problem without transforming the graph. Third, we present the concurrent reduce and peel approach DRP, which combines this heuristic with the metaheuristic Concurrent Difference Core Heuristic [26].

## 5.1 | The Reduce and Transform Routine

We now give an overview of the components of our transformation routine to utilize good-performing independent set solvers to solve the MW2PS problem. An overview is given in Algorithm 1. For a given graph $G$, the main idea of our approach is to build a square graph $G^2$ on which a maximum weight independent set is equivalent to a maximum weight 2-packing set on the original graph, analog to our approach for unweighted 2-packing sets in [9]. We can apply well-studied maximum-weight independent set solvers on this transformed graph to find (near-)optimal solutions. This transformation to the square graph increases the number of edges. Since this results in much denser graphs, this plain transformation becomes slow and requires substantial memory. To alleviate this issue, we add a preprocessing step in which we exhaustively apply our new problem-specific data reductions to obtain a reduced link-graph. Then, we apply the transformation on this instance, resulting in a significantly smaller MWIS instance. The transformation from a link-graph $\mathcal{G} = (G, \mathcal{L})$ is done as follows. First, we compute the square graph $G^2$ of $G$ by adding all edges between vertices in $G$ that are at distance two in $G$. Then, we add all links from the set $\mathcal{L}$ as additional edges in $G$. We apply a maximum weight independent set solver on this transformed instance to obtain an (optimum) solution. Finally, the solution is transformed into an (optimum) solution on the input instance by restoring the reductions. This routine can be used with any maximum weight independent set solvers.

**Algorithm 1** Pseudocode for reduce&transform.

```
input graph G = (V, E, w)
output MW2PS
procedure reduce&transform(G)
  K ←reduce(G)
  K² ←transform(K)
  MWISSolve(K²)
```

## 5.2 | Baseline Reduce-and-Peel Solver redW2pack

The pseudocode for our baseline approach redW2pack is shown in Algorithm 2. This approach alternates between exhaustively applying exact data reductions and a heuristic peeling step.

The algorithm redW2pack *peels* a vertex which is selected by a (randomized) heuristic strategy, i.e. it either *includes* or *excludes* a vertex with respect to a heuristic rating, whenever our exact reduction style `core` was exhaustively applied. This iterative process continues until the graph is empty, and consequently, a heuristic solution for the reduced instance $\mathcal{K}$ is obtained. The peeling step possibly opens up the reduction space so that the exact `core` reductions can *reduce* further vertices.

With reduce-and-peel solvers, a heuristically excluded vertex can lead to a non-maximal solution when the reductions are undone. If possible, we fix this by simply including them when unrolling the stack of applied reductions. Other reductions face the same issue as well. To ensure that the solution for $\mathcal{K}$ is maximal, we maximize the solution greedily by adding free vertices of the largest weight.

Other reduce-and-peel solvers for the MWIS, such as HtWIS, restrict reductions to reduce only vertices of a small degree. They keep track of the vertex degrees to efficiently apply reductions when new small-degree vertices arise from the peeling

**Algorithm 2** Pseudocode for redW2pack.

```
input graph G = (V,E,w), heuristic config C
procedure redW2pack(G, C)
  set_config(C)
  𝒢 ← (V,E,w,∅)
  while 𝒢 not empty:
    𝒢 ←exact_reduce(𝒢)
    𝒢 ←heuristic_reduce(𝒢, p)
```

phase. On the contrary, our approach uses the full set of core reductions and employs the built-in dependency-checking to test reductions only for vertices in regions where the neighborhood has changed.

*Heuristics.* We utilize three heuristic ratings, well-known for the MWIS [28, 29]. We call these ratings weight_diff ($w(v) - w(L(v)) - w(N(v))$), weight ($w(v)$), and degree ($-\deg v - \deg_{\mathrm{L}} v$). The ratings are updated throughout the modifications to the instance using priority queues. We maintain the best $k$ candidates in an array and choose uniformly at random one of them in the peeling step. Since randomness is very crucial for different D-Core instances, we also use a *non-adaptive* approach that pre-computes the rating and does a perturbation with a probability of $p \in [0.5, 1]$ of the remaining ranking before each peeling step. The intuition behind *excluding* a vertex with a small weight_diff rating is that a subset of its neighbors is likely to be part of an optimal solution; a small weight rating suggests that the vertex is unlikely to be part of it; and removing a vertex with a small degree rating possibly entails many new reduction applications. Further, one can *include* a vertex with the highest rating for weight_diff and weight. Our preliminary experiments indicate that if solution quality is highly important, it is wiser to *exclude* a vertex heuristically rather than *including* it. A possible reason is that including a vertex has a wider impact since its neighbors are consequently excluded, while the proposed heuristics only capture very local information.

## 5.3 | Difference-Core Reduce and Peel

We now give the high-level idea of our heuristic Difference-Core Reduce and Peel (DRP), followed by an overview of the main components. A major drawback of the basic reduce-and-peel approaches as redW2pack is that wrong decisions during the heuristic steps cannot be undone. With the new algorithm DRP, we aim to overcome this issue using a new meta-heuristic. We present pseudocode for DRP in Algorithm 3.

*High-level Description.* We propose a new solver called DRP, which uses the baseline reducing-peeling approach redW2pack combined with the meta-heuristic Concurrent Difference Core Heuristic introduced by Großmann *et al.* [26].

The Difference Core (D-Core) is a subgraph of the input instance. It is constructed by first computing multiple solutions for an instance and then removing all vertices that are always included or excluded in all solutions generated. This way, the D-Core contains only the vertices where the heuristic is unsure about the decision, and it is more likely to find improvements. We run the redW2pack algorithm multiple times with different heuristic strategies to generate different solutions.

By running redW2pack multiple times with different heuristic strategies, we can increase the randomization of our approach. Throughout this process, the best solution found is maintained. Furthermore, this approach enables us to return to the original instance without being stuck with a potentially wrong decision while still utilizing the information about the different solutions computed to improve the overall performance.

*The Exact Reduced Instance.* First, an exact reduced instance of $\mathcal{G}$, namely $K$ is obtained by exact_reduce using the strong reduction style introduced in Section 4.2. All subsequent steps are performed directly on this reduced instance $\mathcal{K}$.

*Building the Difference Core.* The D-Core, as introduced by Großmann *et al.* [26] is constructed by using a set of solutions $S = \{S_1, \ldots, S_k\}$ to a given problem. With this set of solutions, the D-Core is defined as the induced link-subgraph of a set of vertices $D \subseteq V$ where the solution status of the vertices in $D$ is different in at least one of the solutions in $S$. Formally, the set $D$ is defined by $D = \{v \in V \mid \exists S_i, S_j \in S : v \in S_i \land v \notin S_j\}$, additionally we defined the set of *similar* vertices as $U = V \setminus D$. Intuitively, the D-Core $G[D]$ contains more difficult parts of the instance. In our approach, the D-Core is an MWIS subproblem yielded by applying our reduce&transform routine to the the link-graph $\mathcal{G}[D]$. When we find a better solution on the D-Core than the best-found solution so far, we embed the solution into the current best solution for $\mathcal{K}$. In the following, we explain how we construct this set of solutions $S$. This approach differs from the original approach by Großmann *et al.* [26].

Generally, when a new best solution is found, we restart with an empty set $S = \emptyset$. Then, we compute different solutions using variations of the redW2pack heuristic until certain conditions for the D-Core are met. Note that the larger the set $S$, the more likely it is that the D-Core is of bigger size. The D-Core instances are expected to be small for a small number of solutions. Therefore, we employ a threshold $\phi$ to restrict the size of the D-Core. We solve the D-Core if the amount of similar vertices relative to the vertices in $\mathcal{K}$ falls below $\phi$. If the solution for the D-Core is already optimal, we slightly decrease $\phi$ to $\phi_-$ times the current measured similarity to observe more redW2pack solutions. On the other hand, if an improvement was made or the solver could not find an improving solution due to a time limit exceeded, we increase $\phi$ by a factor of $\phi_+$ as smaller D-Core instances might be easier to solve. In the latter case, we restart computing the set of solutions $S$ using our redW2pack approach.

*Diversification.* To diversify the solutions computed by redW2pack, we iterate through the heuristic ratings in the order they are introduced above and pick one configuration given the $i$-the step in next_config. In order to solve diverse D-Core instances, we also alternate between the adaptive and the non-adaptive rating (in this order) and even refine these two strategies by alternating between *exclude* and *include* in the case of weight_diff and weight. Whenever an adaptive rating strategy is used, we slightly modify it by incrementing $k$ by 1; if a non-adaptive strategy is re-used, we choose a new $p \in [0.5, 1]$ uniformly at random. The exact reduction phase is diversified by shuffling the order of candidates before a reduction is tested.

*Solving the Difference Core.* To solve the D-Core, we use our reduce&transform routine, which can be combined with any MWIS solver. In this work, we propose two configurations. The first, DRP-BChils uses the baseline local search used in the CHILS heuristic [26], and the other, DRP-KaMIS uses the exact branch and reduce solver KaMIS b&r [37]. We also add a configuration not using the D-Core strategy for comparison.

**Algorithm 3** Pseudocode for DRP.

```
input graph G = (V, E, w), similarity threshold φ, scaling factors φ+ and φ- for the
similarity threshold, maximum time limit for solving the D-CORE t_H
procedure DRP(G, φ, φ+, φ-, t_H)
  K ←exact_reduce(G)
  if n(K) = 0 then return restore(K)                    ▷ restore solution on original graph
  i ← 0
  S ←redW2pack(K, C)
  U ← V_K                                                ▷ similar vertices
  while not time limit exceeded do
    while |U|/n(K) > φ and not time limit exceeded do
      C ←next_config(i):
      S' ←redW2pack(K, C)
      if w(S') > w(S) then S ← S'; U ← V_K                ▷ reset similar vertices
      else U ← U ∩ (V_K \ (S △ S'))                       ▷ update similar vertices
    S_H ←reduce&transform(K[V_K \ U])                     ▷ for t_H · |V_K \ U| / n(K) seconds
    if w(S_H) > w(S \ U) then
      S \ U ∪ S_H ;
      if φ+ · φ < 1 then φ ← φ+ · φ                       ▷ consider smaller cores
    else if S_H is optimal then φ ← φ- · |U|/n(K)         ▷ consider larger cores
    else
      U ← V_K                                            ▷ exact solver timed out
      if φ+ · φ < 1 then φ ← φ+ · φ                       ▷ consider larger cores
return S
```

# 6 | EXPERIMENTAL EVALUATION

**Methodology.** We implemented our algorithm using C++17. The code is compiled using g++ version 12.2 and full optimizations turned on (-O3). To investigate the allocated memory, we use the `malloc_count` library†. We used a machine with an AMD EPYC 7702P (64 cores) processor and 1 TB RAM running Ubuntu 20.04.1. We ran all our experiments with four different random seeds and report geometric mean values unless mentioned otherwise. We set the time limit for all algorithms to 4h and a memory limit of 200 GB. In both cases, we report the best solution found until this point. To compare different algorithms, we use performance profiles [18], which depict the relationship between the objective function size, memory consumption, or running time of each algorithm compared to the best-performing algorithm on the corresponding instance. Each algorithm's performance profile yields a non-decreasing, piecewise constant function. In the following, we denote the number of graphs in the data set with $\#G$. For performance profiles, on the x-axis, we have a factor $\tau$, while the y-axis gives a fraction of the instances (*foi*) of the data set. Specifically, the performance profile for solution quality presents the fraction of instances where the objective function is less than or equal to $\tau$ times the best objective function value found for an instance, i.e. $\#\{\text{objective} \leq \tau * \text{best}\}/\#G$. Here, "objective" refers to the result obtained by an algorithm on an instance, and "best" corresponds to the best result for the same instance among all the algorithms in the comparison. When considering the running time, we present the fraction of instances where the time taken by an algorithm is less than or equal to $\tau$ times the time taken by the fastest algorithm on that instance.

For comparing the running time, the y-axis displays the fraction of instances where the running time taken by an algorithm to find the best solution is less than or equal to $\tau$ times the time taken by the fastest algorithm on that instance, i.e. $\#\{\text{t} \leq \tau * \text{ fastest}\}/\#\text{instances}$. Here, "t" represents the time an algorithm takes on an instance, and "fastest" refers to the time taken by the fastest algorithm on that specific instance. Here, the parameter $\tau$ is greater than or equal to 1. Analog to the running time, for the memory consumption, the profile displays the fraction of instances where the memory consumed by an algorithm is less than or equal to $\tau$ times the memory needed by the most memory-efficient algorithm on that instance.

**Overview/Competing Algorithms.** We perform a wide range of experiments. First, we perform experiments to investigate the influence of the data reduction rules in Section 6.1. We use different MWIS solvers on the square graphs and compare these with the solvers combined with our reduce&transform routine. In particular, we examine the performance of HILS by Nogueira *et al.* [43], HtWIS by Gu *et al.* [29] as well as KaMIS b&r by Lamm *et al.*[37] and m$^2$wis+s by Großmann *et al.* [28]. In the second part of our experiments, we examine the performance of our heuristic algorithm DRP using the two MWIS algorithms KaMIS b&r and CHILS for solving the D-Core instances. Finally, these two configurations are compared to the MWIS solvers combined with reduce&transform in Section 6.4.

**Data Sets.** Our experiments were conducted on 205 graphs, with at least 1,000 vertices. This set of graphs contains 40 SNAP [39], six SSMC [16], five FE [56], 14 MESH [44], and 34 Open Street Map OSM graphs [8, 1] which are frequently used among benchmarks for MWIS solvers [24, 26, 28, 29]. Further, we used 18 large graphs used by Borowitz *et al.* comparing solvers for the Maximum Cardinality 2-Packing Set problem [9]. We assigned weights for these 18 graphs using five different weight distributions (uniform, geometric, hybrid, degree, and unit weights). This results in an additional 90 weighted graphs. The OSM instances are rather small in terms of vertices. Therefore, we added another eight Open Street Map (OMS) graphs from the 10th DiMACS challenge [45], assigning weights sampling from a uniform distribution and using hybrid weights. We give a detailed overview of all graphs and weight distributiuon in Table 1 in the Appendix.

## 6.1 | Impact of Data Reductions on Instances

In this section, we analyze the efficiency of our reductions for preprocessing. First, we compare different configurations of our reduce&transform routine. Then, we compare how different solvers can benefit from our preprocessing routine.

Table 2 presents the performance of reduce&transform with different configurations. Each reduction configuration is defined via the reductions used and their order given in the corresponding reduction lists. In the `full` configuration, we first apply our fast reductions once, followed by our core reductions as presented in Table 1.

The intuition behind the chosen order for the core part is that we want to apply reductions that *include* vertices first since they reduce bigger parts of the graph. The Domination is placed before the simplicial weight transfer reduction and split intersection removal reduction because it is a special case of the latter. If none of the latter reductions are applicable anymore, we try to exclude vertices and fold neighborhoods. To evaluate the impact of our fast data reductions, we split `full` into a `fast` and a `core` configuration, where `fast` only uses the efficient data reductions from Section 4.4 and `core` uses only those from the core part except for the special case of neighborhood folding. Finally, we propose an equally effective but lighter configuration than `full`, called `strong`, that does not use Domination. In Table 1, we give an overview of all configurations compared.

† The `malloc_count` library can be found at https://github.com/bingmann/malloc_count.

| Configuration | Reduction Order | | | | | | | | | | Transformation on | Solver |
|---|---|---|---|---|---|---|---|---|---|---|---|---|
| | fast | | | core | | | | | | | | |
| full | [ 11 | 12 | 13 | 2 | 7 | 9 | 6 | 4 | 10 ] | | reduced link-graph | any MWIS sovler |
| fast | [ 11 | 12 | 13 | | | | | | ] | | reduced link-graph | any MWIS solver |
| strong | [ 11 | 12 | 13 | 2 | | 9 | 6 | 4 | 10 ] | | reduced link-graph | any MWIS sovler |
| core | [ | | | 2 | 7 | 9 | 6 | 4 | 10 ] | | reduced link-graph | any MWIS sovler |
| transform | [ | | | | | | | | ] | | full graph | any MWIS solver |

**TABLE 1** Overview of all reduce&transform configurations evaluated in our experiments.

In Table 2, we report the sizes of the transformed graphs and the time and memory needed to compute them. Furthermore, we give the number of fully reduced graphs, i.e. the number of instances solved to optimality by our reductions, as well as the geometric mean reduction weight offset. To emphasize the efficiency of our proposed reduce&transform configurations, we compare them with the configuration `transform` that determines the square graph by building every 2-neighborhood. For an overview of all configurations compared see Table 2.

Figure 6 shows that using `fast` is always better than directly transforming the graph in all aspects regarding running time, graph sizes, and memory consumption. With this `fast` configuration, we reduce the instances on average to less than 50 % of the original number of vertices while saving up to 39 % of memory and using less time. The results for the `core` configuration clearly show the importance of our fast reductions. Compared to the `full` configuration, we can improve the running time by approximately a factor of 4 while reducing the memory consumption by 27 %. The `strong` configuration is the best performing regarding the size of the reduced transformed graph as well as memory consumption. Furthermore, on average, it uses only a factor of 1.88 more time than our fastest variant.

Figure 7 gives a more detailed insight into the reduction impact for our configurations regarding the reduced instance size: with our core reductions, we find for almost all instances a reduced instance with at most 25 % vertices and edges relative to the square graph. However, there are outliers left where our reductions are hardly applicable. Regarding the configuration `fast`, we observe that its effectiveness varies highly. The first quantile is approximately at 24 % (5 %) remaining vertices (edges), the median at 50 % (27 %), and the third quantile at 90 % (95 %).

The configurations utilizing our core reductions can solve at least 43 % of the instances to optimality. Although `fast` is not capable of reducing any instance fully, we still want to point out that when using fast and core reductions in conjunction (`strong` or `full`), there are instances where we can reduce running times from almost 3 hours down to less than 10 seconds as observed for texttsnap_wiki-Talk-uniform. This graph has almost 74 % degree-one vertices, which are efficiently reduced with our Fast Degree-One Reduction, whereas `core` and `transform` are unable to find a (reduced) and transformed instance.

| | median [%] | | geo. mean | | | |
|---|---|---|---|---|---|---|
| Configuration | $n(\mathcal{K})/n(G^2)$ | $m(\mathcal{K})/m(G^2)$ | $t$ [$s$] | mem. [MB] | $o(\mathcal{K})$ | #opt [%] |
| core | **0.1** | **0.000 2** | 1.38 | 89.8 | 312 958 | **44** |
| strong | **0.1** | **0.000 2** | 0.34 | **61.8** | 312 962 | **44** |
| fast | 50.0 | 26.372 5 | **0.18** | 61.9 | 126 430 | - |
| full | **0.1** | **0.000 2** | 0.37 | 65.3 | **312 963** | **44** |
| transform | 100.0 | 100.000 0 | 0.30 | 101.7 | - | - |

**TABLE 2** Performance of reduce&transform with different configurations. The set is restricted such that all configurations can create the transformed graph. The geometric mean offset $o(\mathcal{K})$ was computed only for instances where all variants found an offset larger than zero.

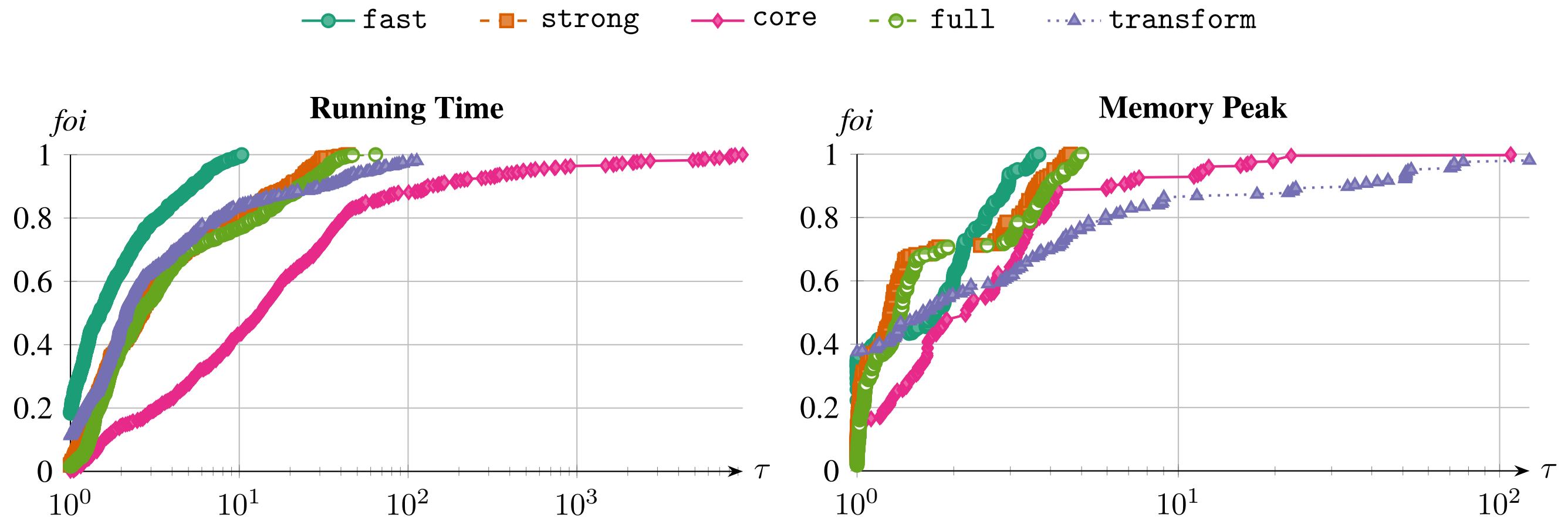

**FIGURE 6** Performance profiles comparing different configurations of reduce&transform, defined in Table 1. The left figure shows the running time to compute the (reduced) transformed graph, and the right figure presents the memory peak. The y-axis shows the corresponding fraction of instances (*foi*) solved faster or with less memory than $\tau$ times the best-performing algorithm on the respective instance.

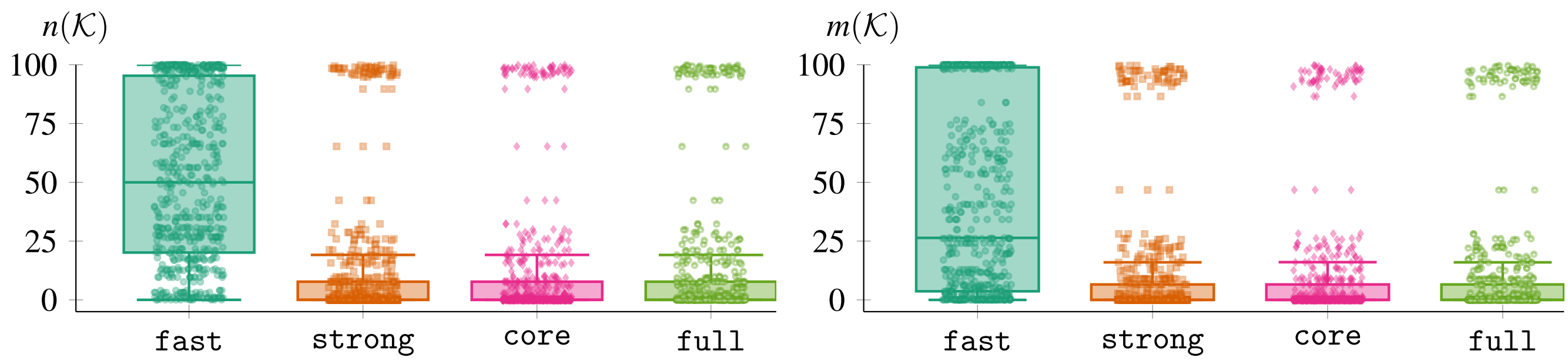

**FIGURE 7** Comparison of reduced instances $\mathcal{K}$ computed by different reduction configurations defined in Table 1. On the right, we present the remaining number of vertices and on the left the remaining number of edges in % relative to the square graph.

The performance profiles regarding the running time and memory peak in Figure 6 provide revealing results. In terms of running time, `strong`, `full`, and `transform` perform very similarly. Almost 50 % of the runs are at most a factor of 2 slower than the fastest running times among all configurations on the respective instances. For 20 % of the runs, on the other hand, Figure 6 indicates that these are at least a factor of 10 up to 100 slower than the respective fastest running times. In the case of `fast`, we notice that it is at most a factor of 10 slower on all runs while having the best running times on 20 % of the instances. The configuration `core` performs worst regarding running time. It is up to a factor of $10^4$ slower than the fastest configuration. However, `core` fully reduces 43 % of the instances while `transform` only determines the square graph. In this sense, it is much to our surprise that `strong` and `full` can keep up with the performance of `transform` while they are at least as good in solution quality as `core`. The performance profiles regarding the memory peak clearly indicate that using our reductions is better than directly transforming the graph. Without reductions, `transform` needs up to a factor of $10^2$ more memory than the smallest achieved peaks, while our reduce&transform configurations with(out) fast reductions are at most a factor 5 (20) apart from the smallest peaks.

Our experimental results underline the efficiency in terms of running time and memory consumption for `fast`, `strong`, and `full` gained by using our fast reductions. Combining this preprocessing phase with our core reductions results in an effective reducing scheme that yields for 44 % of the instances optimal solutions, with median reduced instance sizes of $0.1$ % vertices and $0.0002$ % edges relative to the square graph. Moreover, `transform` requires multiple orders of magnitude more memory and takes a similar amount of running time, making it inapplicable on some instances under reasonable resource limitations. Regarding the `full` configuration, it appears that the rule Domination has only a small positive impact compared to `strong`.

| MWIS solver | Configuration | $w(S)$ | $t$ | mem. | opt. | reached opt. | # add. sol. |
|---|---|---|---|---|---|---|---|
| HILS | `strong` | 211 321 | 0.943 | 33.6 | 43.9 | 98.6 | 31 |
| | `fast` | 211 222 | 20.291 | 39.2 | 0.0 | 74.6 | 15 |
| | `transform` | 211 101 | 171.181 | 75.8 | 0.0 | 62.0 | 8 |
| CHILS | `strong` | **211 445** | 0.722 | 38.6 | 43.9 | 99.3 | 30 |
| | `fast` | 211 438 | 23.638 | 61.0 | 0.0 | 67.6 | 30 |
| | `transform` | 211 436 | 73.687 | 147.9 | 0.0 | 72.1 | 24 |
| HtWIS | `strong` | 210 582 | **0.168** | **33.5** | 43.9 | 76.8 | **32** |
| | `fast` | 210 437 | 0.171 | 37.9 | 32.2 | 65.5 | **32** |
| | `transform` | 210 418 | 1.107 | 70.0 | 29.3 | 61.3 | 26 |
| m$^2$wis | `strong` | 211 296 | 0.889 | 45.3 | 59.5 | **100.0** | 26 |
| | `fast` | 211 304 | 1.069 | 69.4 | 15.6 | **100.0** | 30 |
| | `transform` | 211 299 | 5.310 | 205.3 | 15.1 | 98.6 | 22 |
| KaMIS b&r | `strong` | 209 822 | 0.369 | 50.4 | **69.3** | **100.0** | 30 |
| | `fast` | 209 744 | 0.318 | 77.9 | 68.8 | 99.5 | 30 |
| | `transform` | 209 742 | 0.767 | 225.4 | 67.8 | 97.9 | 26 |

**TABLE** 3 Performance of our reduction configurations in combination with various MWIS solvers. The columns $w(\mathcal{S})$, $t$ and mem. show the geometric mean solution quality (including reduction offset), time found (in seconds, including reduce&transform time), and memory peak (in MB, including reduce&transform memory). These are evaluated on the set of instances where all solvers found a 2-packing set. The 'opt.' column shows the percentage of solutions proven optimal by the respective solver. The column 'reached opt.' shows the percentage of instances solved optimally among the instances where at least one solver proved optimality. The last column shows the number of feasible solutions found in addition to the common set of solutions found by the respective solver. The **best** results are highlighted in bold.

**Observation for Reduction Impact on Instances**
Considering the impact of reductions on the (reduced) transformed instance, our `fast` configuration is always preferable to `transform`. It is faster while requiring less memory than `transform`. The configuration `strong` needs, on average, only 0.002 seconds more than `transform`, but uses the least amount of memory and already solves 44 % of our instances optimally.

## 6.2 | Impact of Data Reductions for Solving

In the following, we investigate the practical effect of the configurations `fast`, `strong`, and `transform` when solving the transformed graph with a wide range of MWIS solvers. All running times and memory presented in this section always include the time and memory used for reduce&transform and the solving process combined. For the solution qualities, we always give the reduction offset and the solution quality found on the reduced and transformed instance found by the different solvers. Besides the exact solver KaMIS b&r, we add four heuristics in this comparison. We include the evolution-based solver m$^2$wis+s and a reduce and peel solver HtWIS. Additionally, we have the local search algorithm HILS and the concurrent local search CHILS. In Table 3, we compare the different reduction configurations and solvers. We present the geometric mean running time and solution quality. Moreover, we also report the number of additional instances solvable through our preprocessing, mainly due to memory issues with the original square graph. In Figure 8, we compare the different reduction configurations for each solver in more detail.

Table 3 shows that all algorithms benefit from using our preprocessing instead of the plain graph transformation. When comparing `transform` and `fast`, we see speedups of more than a factor of 8 (HILS) while still improving on solution quality. Additionally, the amount of memory needed is reduced by almost a factor of two. These factors are further increased when comparing `transform` to the `strong` variant. Here, we see speedups up to a factor of 180 (HILS) using even less memory than `fast` and further improving on solution quality. The algorithm m$^2$wis+s is the only one not benefiting from the `strong` variant improvements in solution quality. However, the `fast` variant can improve running time, solution quality, and memory used. In this heuristic, additional independent set reductions are included, which is why it can increase the percentage of instances proven to be optimal from 43.9 %, which are the instances fully reduced by our `strong` reductions, to 59.5 %. The fastest and most memory-efficient configuration is `strong` combined with HtWIS. This configuration has the most additional feasible solutions. However, this comes with a loss in solution quality compared to the other heuristics, which we see by the small percentage of only 76.8 % of instances where it reached a (proven) optimal solution. In contrast to HtWIS, all other `strong` configurations reach 98.6 % or higher.

| | Configuration Details | | | | | Results | | |
|---|---|---|---|---|---|---|---|---|
| Configuration | core solver | $t_H$ | $\phi$ | $\phi_+$ | $\phi_-$ | $w(S)$ | $t$ | mem. |
| DRP-BChils | CHILS Baseline | 80 | 0.6 | 1.00 | 1.00 | **446 750** | 4.3 | 72.0 |
| DRP-KaMIS | KaMIS b&r | 80 | 0.8 | 1.05 | 0.95 | 446 687 | **2.8** | 78.9 |
| DRP-no-core | - | - | - | - | - | 445 710 | 3.8 | **71.6** |

**TABLE 4** Geometric mean results of solution weight $w(S)$, time found $t$ (in seconds), and memory peak mem. (in MB) for the different configurations of our algorithm DRP. The last configuration failed to finish solving for one instance within the set timelimit. All configuration are evaluated on the set of instances where all algorithms found a solution in time.

Regarding KaMIS b&r, we note that it often times out. In case of a timeout, the branch and reduce solver greedily builds a maximal solution for the remaining non-solved connected components.

In Figure 8, we see that for *all* solvers, the basic `transform` is worse concerning solution quality, running time, and memory peak compared to both reduction configurations `fast` and `strong`. With `strong`, the memory peak compared to `transform` for all solvers tested on around 20 % of the instances is reduced by more than one order of magnitude. Especially for the local search algorithms, HILS and CHILS, we see an additional huge improvement in the running time when comparing `strong` with `fast`. This speedup is up to multiple orders of magnitude, while the `strong` configurations are also further improving solution quality and reducing the memory peak observed.

**Observation for Reduction Impact on Solving**
When considering the performance of different solvers on the (reduced) transformed instances, while always adding the transformation time to the solve time, we see that the time needed for the `strong` configuration is well spent, and overall, these configurations find better solutions multiple magnitudes faster with less memory needed. Especially local search heuristics benefit from using the `strong` configuration.

## 6.3 | The Algorithm redW2pack

In this set of experiments, we investigate the different configurations of our solver redW2pack. We present the parameters for the different configurations in Table 4. The parameters chosen worked best in preliminary experiments. Configurations with values for $\phi_-$ and $\phi_+$ work well if they change $\phi$ only slightly. The intuition behind that is that if $\phi_-$ decreases $\phi$ too strong, it might need many new solutions to compute a large D-Core, which in the end might be hard to solve optimally. A good choice for the initial $\phi$ tends to yield D-Core instances of manageable size that the MWIS solver can solve in a reasonable time. Additionally to the parameters, we present the geometric mean results for different configurations of solution quality and time found in Table 4.

We compare solving the D-Core with the exact solver KaMIS b&r resulting in the configuration DRP-KaMIS and the heuristic baseline local search used in the CHILS algorithm, yielding DRP-BChils. Additionally, we present results without using the D-Core strategy with the variant DRP-no-core. Table 4 shows the configurations with the parameter values and the geometric mean solution quality, time found, and memory peak achieved by the configurations.

In Figure 9, we present performance profiles comparing the solution quality, time found, and memory peak for the different DRP configurations in more detail. Note that 44 % of the instances are fully reduced by the `strong` preprocessing and thereby solved optimally for all the configurations compared here. The two variants using the D-Core perform better regarding solution quality. The configuration DRP-no-core computes on more than 30 % of the instances a worse solution than the other configurations. The configuration DRP-BChils performes best in terms of solution quality, while the configuration DRP-KaMIS has the fastest times to find the best solutions overall. It is counterintuitive that the configuration DRP-BChils is the best regarding solution quality, as the exact solver KaMIS b&r should be able to find better or equal solutions. However, the exact solver KaMIS b&r cannot solve all D-Core instances to optimality within the time limit of 80 seconds. Since the baseline local search in CHILS is a very powerful algorithm, it computes near-optimal solutions fast. This explains why the configuration DRP-BChils is better regarding solution quality. Nevertheless, the local search approach always utilizes the full time limit to solve the D-Core, resulting in worse performance regarding running time. On the other hand, the variant using KaMIS b&r can save time when computing the D-Core solutions, as it continues as soon as the solution found is proven optimal. The configuration DRP-BChils is almost as memory efficient as the configuration DRP-no-core, while the approach using the exact solver DRP-KaMIS needs up to 4 times as much memory.

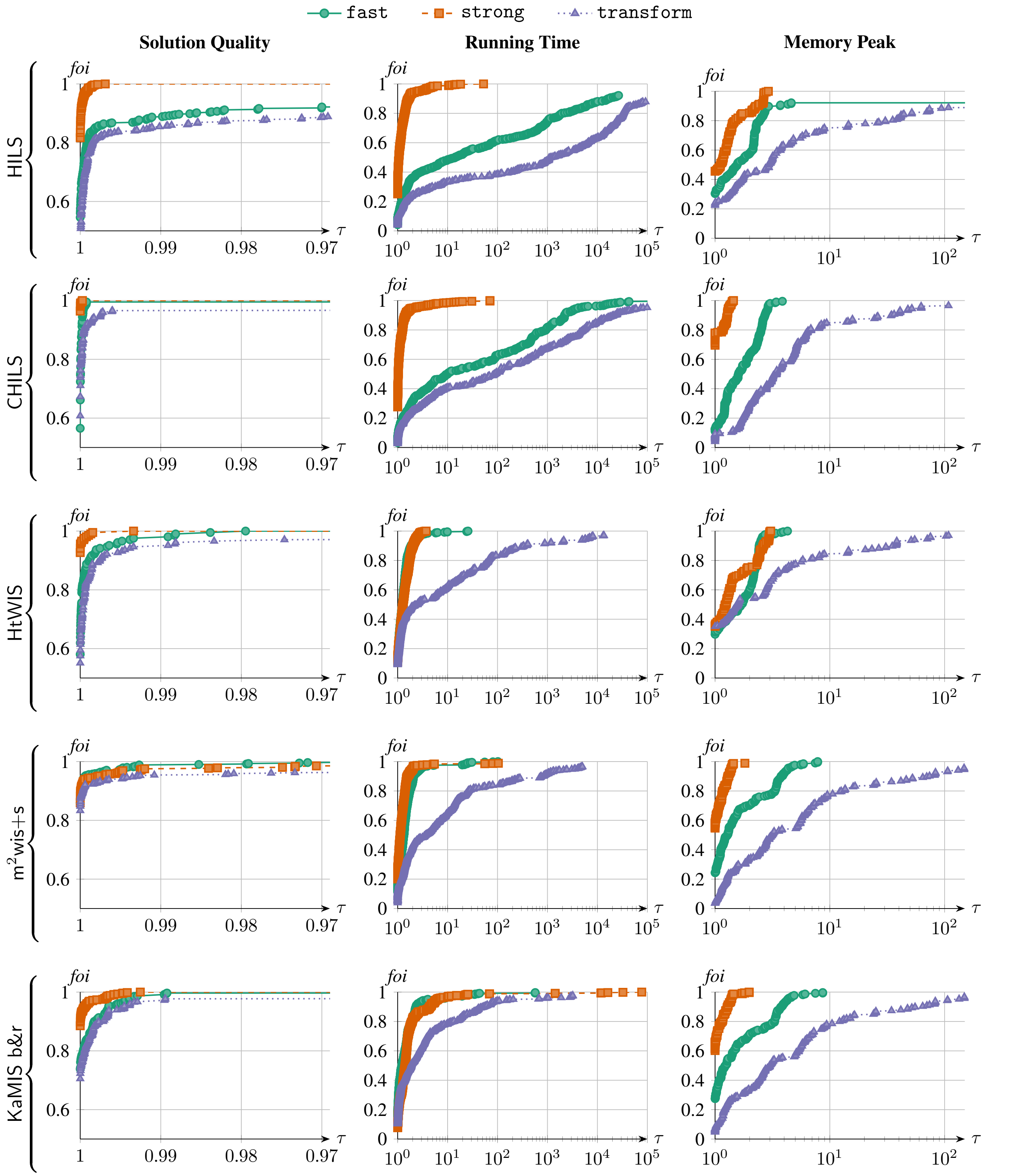

**FIGURE 8** Performance profiles for reduction configuration comparison using different MWIS solvers (different rows). Note that the y-axes—showing the fraction of instances (*foi*)—starts at 0.5 for the solution quality. We present profiles for solution quality (left), running time to find the best solution (center) and the memory consumption (right).

**Observation for DRP Configurations**
Our experiments for the different DRP configurations clearly show that using the D-Core strategy can be beneficial for improving the solution quality and also running time. When using DRP-KaMIS, the algorithm can find the best solutions on average a factor of 1.4 faster than DRP-no-core. On the other hand, the configuration DRP-BChils yields the best solution quality on average while being almost as memory efficient as DRP-no-core.

## 6.4 | State-of-the-Art Comparison

Figure 10 shows performance profiles comparing the best-performing configurations for all solvers on all instances. We present comparisons of solution quality, running time, and memory peak. Table 5 summarizes these results for the different solvers.

We can observe that the configuration strong-CHILS performs overall the best regarding solution quality, while the DRP configurations are very close to this performance. The configuration strong-HtWIS is the fastest configuration with the overall lowest memory peak, but it performs worst regarding solution quality. The highest memory peak is observed for the configuration fast-m$^2$wis. However, this is the only configuration that uses the fast reduction routine, which means initially a bigger transformed reduced graph compared to the other approaches.

A detailed comparison of the different approaches on interesting instances is presented in Table 6. Here, we present per instance results for the 40 largest instances where the performance of the algorithms differed most regarding either solution quality or running time. Looking at different graph classes, we see that fast-m$^2$wis performs particularly well for different osm instances. This observation can be explained by the fact that m$^2$wis has several reductions for the MWIS implemented, which work well on these OSM instances. For instances where these reductions do not work well, e.g. snap or mesh, the strong-CHILS approach excels. Furthermore, the different weight distributions for the instances do not significantly affect the difference in performance between the algorithms. For example, for every weight distribution, the instance *road_usa*—which is the second largest instance in our data set—is always solved best by DRP-BChils.

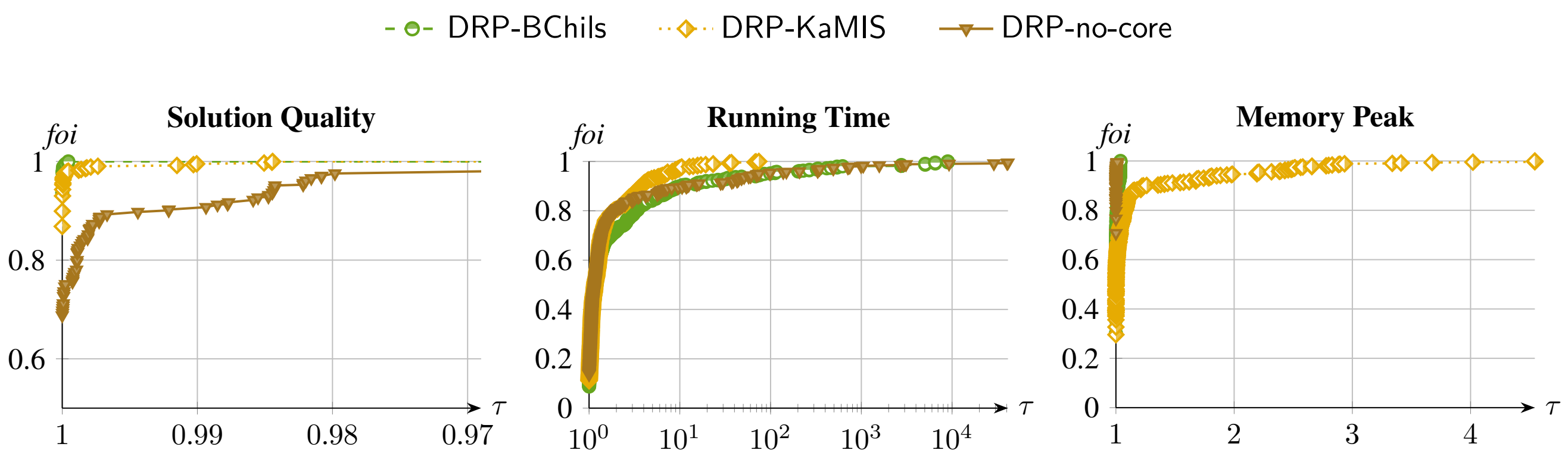

**FIGURE 9** Performance profiles comparing the different configurations of DRP regarding solution qualtiy (left), time for the best solution to be found (center) and the memory peak (right).

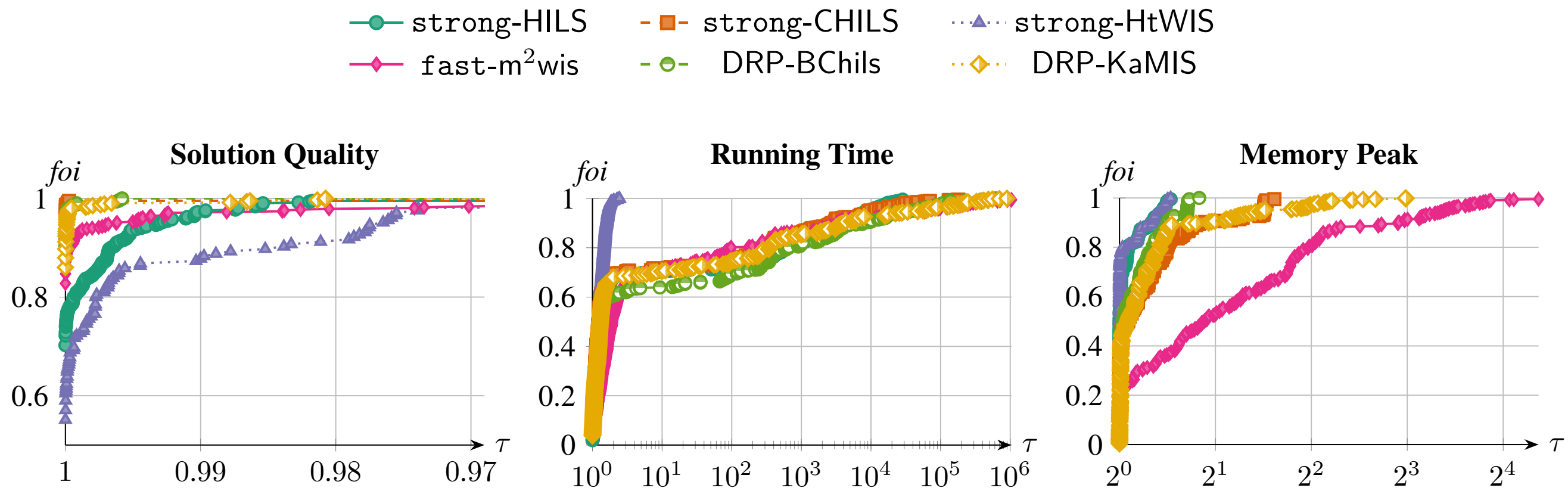

**FIGURE 10** Performance profiles for the state-of-the-art comparison. Configurations of best performing reduce&transform configurations and DRP configurations in terms of solution quality.

| Solver | $w(S)$ | $t$ | mem. | opt. | reached opt. | # add. sol. |
|---|---|---|---|---|---|---|
| strong-HILS | 441 458 | 2.88 | 68 | **44** | **100** | 3 |
| strong-CHILS | **441 885** | 2.09 | 78 | **44** | 99 | 2 |
| strong-HtWIS | 440 202 | **0.34** | **67** | **44** | 85 | **4** |
| fast-m$^2$wis+s | 441 454 | 2.89 | 140 | 16 | **100** | 2 |
| DRP-BChils | 441 864 | 4.36 | 72 | **44** | **100** | **4** |
| DRP-KaMIS | 441 801 | 2.82 | 79 | **44** | **100** | **4** |

**TABLE 5** Performance of our reduction configurations in combination with various MWIS solvers. The columns $w(\mathcal{S})$, $t$ and mem. show the geometric mean solution quality, time found (in seconds) and memeory peak (in MB). These are evaluated on the set of instance where all solvers found a 2-packing set. Note that in contrast to Table 3, these are evaluated on more instances. The 'opt.' column shows the percentage of solutions that were proven to be optimal by the respective solver. The column 'reached opt.' shows the percentage of instances that were solved optimally among the set of instances where at least one solver proved optimality. The last column shows the number of further feasible solutions found in addition to the common set of solutions found by the respective solver. The **best** results are highlighted in bold.

**Observation for State-of-the-Art Comparison**
In the comparison between the different DRP configurations and reduce&transform combined with different state-of-the-art MWIS solvers, we can see that both DRP configurations can keep up with the state-of-the-art MWIS solvers combined with reduce&transform. Additionally, on some of the biggest instances (*europe* and *road_usa*), DRP-BChils and DRP-KaMIS can find the best solution quality, outperforming all MWIS approaches.

## 7 | CONCLUSION AND FUTURE WORK

In this work, we are the first to introduce data reduction rules for the MW2PS problem. We use these, in total 13 reductions, to develop a preprocessing routine reduce&transform. This routine reduces and transforms an instance of the MW2PS into an instance for the MWIS problem, such that the solution to the MWIS problem on the transformed graph is an MW2PS on the original graph. With reduce&transform, well-studied, scalable MWIS solvers can solve the MW2PS problem efficiently.

Our experiments show that our reductions can fully reduce and thereby optimally solve 44 % of the instances in our data set. Furthermore, reduce&transform improves solution quality, running time, and memory consumption compared to a naive transformation for every MWIS solver tested. With reduce&transform we achieve speedups compared to a naive transformation up to multiple orders of magnitude.

Moreover, we propose a new heuristic DRP, a reduce and peel approach based on the metaheuristic Concurrent Difference Core Heuristic introduced by Großmann *et al.* [26]. Especially for large graphs DRP excels. Our experiments indicate that our heuristic is able to keep up with the state-of-the-art MWIS solvers equipped with our preprocessing routine reduce&transform. Additionally, DRP can find the best solution quality on the biggest instances in our data set, outperforming all MWIS approaches.

For future work, we want to extend the set of reductions for the MW2PS further. Additionally, since the local search approaches for the MWIS work best, we are interested in engineering a new local search heuristic for the MW2PS to circumvent any graph transformation, thereby saving even more running time and memory.

| | Instance | | DRP-BChils | | DRP-KaMIS | | strong-CHILS | | strong-HILS | | strong-HtWIS | | fast-m$^2$wis+s | |
|---|---|---|---|---|---|---|---|---|---|---|---|---|---|---|
| | | | *w* | *t* | *w* | *t* | *w* | *t* | *w* | *t* | *w* | *t* | *w* | *t* |
| mesh | buddha | uf | 32 602 416 | 14 135.22 | 32 602 593 | 14 203.88 | **32 603 831** | 13 875.78 | 32 021 963 | 14 393.57 | 31 880 965 | **4.57** | 31 447 350 | 14 391.90 |
| mesh | dragonsub | uf | 18 464 865 | 14 199.63 | 18 464 623 | 14 355.58 | **18 466 821** | 13 023.43 | 18 324 835 | 14 395.13 | 17 987 255 | **1.89** | 17 499 416 | - |
| mesh | ecat | uf | 20 963 870 | 14 250.79 | 20 963 796 | 13 993.82 | **20 965 446** | 13 288.50 | 20 773 748 | 14 393.50 | 20 452 008 | **5.61** | 19 600 785 | 14 391.57 |
| osm | asia | h | 104 862 636 | 13 866.77 | **104 862 684** | 14 276.46 | 104 862 634 | 13 853.99 | 104 244 901 | 14 396.80 | 104 629 705 | **23.49** | 104 860 823 | 13 520.92 |
| osm | asia | uf | 124 441 135 | 8 053.17 | 124 441 141 | 8 195.60 | 124 441 137 | 4 428.06 | 124 425 250 | 14 384.99 | 124 403 501 | **17.49** | **124 441 143** | 222.29 |
| osm | belgium | h | 12 625 769 | 9 686.40 | 12 625 796 | 7 404.59 | 12 625 751 | 11 272.41 | 12 607 633 | 14 381.66 | 12 581 266 | **3.52** | **12 625 797** | 202.48 |
| osm | belgium | uf | **14 853 069** | 1 341.68 | **14 853 069** | 46.18 | **14 853 069** | 162.34 | 14 852 592 | 2 039.53 | 14 851 095 | **2.38** | **14 853 069** | 6.22 |
| osm | europe | h | 448 902 095 | 14 250.24 | **448 903 472** | 14 343.86 | 448 899 155 | 14 262.91 | - | - | 447 920 680 | **154.00** | 448 883 604 | - |
| osm | europe | uf | 529 717 578 | 10 019.66 | 529 717 576 | 10 906.18 | 529 717 579 | 10 408.97 | 529 286 469 | 14 397.53 | 529 637 587 | **110.99** | **529 717 599** | 562.93 |
| osm | germany | uf | **119 677 619** | 523.92 | **119 677 619** | 233.19 | **119 677 619** | 314.74 | 119 674 575 | 14 194.11 | 119 665 411 | **25.68** | **119 677 619** | 51.24 |
| osm | germany | h | 102 376 273 | 13 735.38 | 102 376 346 | 12 480.37 | 102 375 878 | 14 196.11 | 101 390 183 | 14 397.53 | 102 139 965 | **30.19** | **102 376 351** | 1 630.99 |
| osm | great-britain | h | 69 224 797 | 11 081.76 | 69 224 825 | 7 045.04 | 69 224 786 | 12 173.27 | 69 018 599 | 14 397.00 | 69 115 568 | **15.57** | **69 224 827** | 551.96 |
| osm | great-britain | uf | **81 011 319** | 423.07 | **81 011 319** | 62.40 | **81 011 319** | 36.63 | 81 010 241 | 10 312.23 | 81 005 926 | **12.05** | **81 011 319** | 25.25 |
| osm | italy | h | 58 703 499 | 12 351.94 | 58 703 552 | 12 163.02 | 58 703 514 | 12 189.68 | 58 547 438 | 14 397.23 | 58 592 323 | **13.10** | **58 703 583** | 6 990.41 |
| osm | italy | uf | **69 970 692** | 579.19 | **69 970 692** | 859.43 | **69 970 692** | 168.25 | 69 968 413 | 13 695.05 | 69 960 840 | **9.72** | **69 970 692** | 35.02 |
| osm | netherlands | h | 19 781 545 | 10 011.18 | 19 781 556 | 3 876.05 | 19 781 549 | 9 415.14 | 19 731 383 | 14 395.07 | 19 718 874 | **4.66** | **19 781 557** | 314.71 |
| osm | netherlands | uf | **22 599 881** | 566.92 | **22 599 881** | 79.47 | **22 599 881** | 8 567.77 | 22 598 638 | 11 823.36 | 22 593 996 | **4.33** | **22 599 881** | 16.92 |
| red2pack | road_central | d | 13 499 922 | 14 179.00 | 13 499 758 | 14 063.58 | **13 499 960** | 14 238.69 | 13 457 810 | 14 395.20 | 13 470 102 | **35.75** | 13 492 263 | 14 392.47 |
| red2pack | road_central | g | **235 169 677** | 14 395.97 | 235 161 329 | 14 342.85 | 235 165 168 | 14 308.29 | 232 195 767 | 14 397.57 | 234 241 460 | **54.75** | 234 484 588 | 14 393.77 |
| red2pack | road_central | h | 549 590 890 | 14 148.78 | 549 590 464 | 14 075.05 | **549 591 087** | 13 750.86 | 547 901 981 | 14 397.23 | 548 981 006 | **52.93** | 549 582 921 | 12 895.80 |
| red2pack | road_central | u | 8 999 961 | 13 959.77 | 8 999 827 | 14 161.04 | **8 999 975** | 14 013.27 | 8 971 625 | 14 396.07 | 8 980 068 | **39.51** | 8 994 753 | 14 392.30 |
| red2pack | road_central | uf | **8 999 967** | 13 923.06 | 8 999 843 | 13 553.04 | **8 999 967** | 14 262.95 | 8 971 109 | 14 397.33 | 8 980 068 | **39.49** | 8 995 065 | 14 392.30 |
| red2pack | road_usa | d | **22 946 461** | 14 270.94 | 22 946 188 | 14 249.10 | 22 946 259 | 14 276.51 | 22 835 809 | 14 396.93 | 22 894 944 | **40.14** | 22 924 695 | 14 392.93 |
| red2pack | road_usa | g | **399 716 160** | 14 396.80 | 399 702 024 | 14 379.61 | 399 705 895 | 14 332.03 | 394 096 678 | 14 395.93 | 398 129 609 | **58.87** | 397 709 419 | 14 397.27 |
| red2pack | road_usa | h | **909 315 888** | 14 298.16 | 909 311 627 | 14 386.40 | 909 311 190 | 14 289.75 | 902 975 074 | 14 396.27 | 907 736 821 | **59.81** | 909 070 721 | 14 394.97 |
| red2pack | road_usa | u | **15 297 660** | 14 263.56 | 15 297 438 | 14 274.22 | 15 297 510 | 14 295.70 | 15 225 993 | 14 397.23 | 15 263 296 | **41.75** | 15 283 063 | 14 393.43 |
| red2pack | road_usa | uf | **15 297 632** | 14 338.10 | 15 297 465 | 14 262.34 | 15 297 512 | 14 289.57 | 15 224 282 | 14 396.47 | 15 263 296 | **41.55** | 15 282 400 | 14 393.13 |
| snap | as-skitter | uf | **22 747 281** | 309.97 | **22 747 281** | 1 944.01 | **22 747 281** | **24.06** | **22 747 281** | 58.17 | 22 745 994 | 27.45 | **22 747 281** | 233.67 |
| snap | com-amazon | f | 6 784 132 | 1 345.69 | **6 784 135** | 673.02 | **6 784 135** | 161.71 | 6 783 500 | 528.77 | 6 779 497 | **1.89** | **6 784 135** | 5.65 |
| snap | roadNet-CA | uf | 69 533 118 | 13 506.27 | 69 533 078 | 13 547.56 | **69 533 187** | 7 500.36 | 69 441 335 | 14 392.67 | 69 298 011 | **6.07** | 69 531 967 | 7 944.87 |
| snap | roadNet-PA | f | 38 549 128 | 10 606.64 | 38 549 112 | 12 495.52 | **38 549 235** | 4 182.21 | 38 526 182 | 14 374.67 | 38 428 331 | **3.17** | 38 549 060 | 8 404.16 |
| snap | roadNet-PA | uf | 38 606 226 | 7 430.02 | 38 606 192 | 13 674.26 | **38 606 300** | 5 360.57 | 38 584 253 | 14 307.96 | 38 483 203 | **2.70** | 38 606 010 | 4 720.88 |
| snap | roadNet-TX | uf | 49 698 766 | 13 048.85 | 49 698 796 | 13 072.83 | **49 698 835** | 11 418.44 | 49 671 265 | 14 357.58 | 49 557 120 | **3.87** | 49 698 111 | 8 575.76 |
| snap | soc-pokec-rel. | uf | 24 557 118 | 7 946.46 | **24 557 259** | 394.37 | 24 557 109 | 6 862.02 | 24 552 172 | 12 212.59 | 24 541 657 | **25.58** | 24 557 199 | 4 671.27 |
| snap | web-BerkStan | f | 3 993 862 | 1 121.93 | **3 993 879** | 415.52 | **3 993 879** | 337.15 | 3 993 828 | 423.79 | 3 992 470 | **158.42** | **3 993 879** | 13 022.76 |
| snap | web-BerkStan | uf | 3 997 024 | 682.55 | **3 997 026** | 238.33 | 3 996 974 | 358.62 | 3 996 962 | 319.13 | 3 995 734 | **210.57** | **3 997 026** | 12 291.58 |
| snap | web-Google | uf | **10 993 272** | 39.57 | **10 993 272** | 2.68 | **10 993 272** | **2.57** | **10 993 272** | 2.82 | 10 993 265 | 3.10 | **10 993 272** | 5.59 |
| ssmc | ca2010 | f | 9 098 481 | 8 025.21 | 9 098 171 | 14 083.31 | **9 098 531** | 2 094.41 | 9 091 051 | 13 683.52 | 9 010 605 | **4.75** | 9 097 549 | 13 037.78 |
| ssmc | fl2010 | f | 4 655 766 | 8 275.04 | 4 655 665 | 12 814.44 | **4 655 789** | 1 009.34 | 4 654 381 | 12 920.20 | 4 635 302 | **2.52** | 4 655 757 | 10 420.59 |
| ssmc | il2010 | f | 3 231 622 | 11 699.56 | 3 231 436 | 13 985.54 | **3 231 632** | 775.99 | 3 228 123 | 14 182.51 | 3 192 156 | **3.46** | 3 231 242 | 10 609.37 |

**TABLE 6** Detailed results for state-of-the-art comparison on "interesting" instances. We present results for the 40 largest instances (in terms of vertices) where at least one solver has a worse solution quality than a factor 0.98 or needs at least a factor 100 more running time than is needed on average for computing the best solution on that instance. If a timeout occurs, we mark it with "-". If no feasible solution was computed until the timeout, we also mark this with "-" in the respective solution column. The instances are grouped by their graph class and have different weight assignments, being uniform (uf), unit (u), hybrid (h), geometric (g), degree (d), or from the file (f).

## REFERENCES

1. *Openstreetmap*, https://www.openstreetmap.org .
2. F. N. Abu-Khzam, S. Lamm, M. Mnich, A. Noe, C. Schulz, and D. Strash, *Recent advances in practical data reduction*, H. Bast, C. Korzen, U. Meyer, and M. Penschuck (eds.), *Algorithms for Big Data: DFG Priority Program 1736*, Springer Nature Switzerland, Cham, 2022. 97–133, .
3. T. Akiba and Y. Iwata, *Branch-and-reduce exponential/FPT algorithms in practice: A case study of vertex cover*, Theoretical Computer Science **609, Part 1** (2016), 211–225.
4. Y. Atsuta and S. Takahashi, *The maximum weighted k distance-d independent set problem on interval graph*, *2020 9th International Congress on Advanced Applied Informatics (IIAI-AAI)*, 2020, 840–841, .
5. D. Bacciu, A. Conte, and F. Landolfi, *Generalizing downsampling from regular data to graphs* (2022).
6. G. Bacsó, D. Marx, and Z. Tuza, *H-Free Graphs, Independent Sets, and Subexponential-Time Algorithms*, LIPIcs, Volume 63, IPEC 2016 **63** (2017), 3:1–3:12.
7. B. S. Baker, *Approximation algorithms for np-complete problems on planar graphs*, J. ACM **41** (1994), no. 1, 153–180. URL https://doi.org/10.1145/174644.174650.
8. L. Barth, B. Niedermann, M. Nöllenburg, and D. Strash, *Temporal map labeling: A new unified framework with experiments*, *Proceedings of the 24th ACM SIGSPATIAL International Conference on Advances in Geographic Information Systems*, 2016, 1–10.
9. J. Borowitz, E. Großmann, C. Schulz, and D. Schweisgut, *Finding optimal 2-packing sets on arbitrary graphs at scale*, CoRR **abs/2308.15515** (2023). URL https://doi.org/10.48550/arXiv.2308.15515.
10. J. D. Chandler, W. J. Desormeaux, T. W. Haynes, and S. T. Hedetniemi, *Neighborhood-restricted [ $\leq 2$ ] -achromatic colorings*, Discrete Applied Mathematics **207** (2016), 39–44.
11. L. Chang, *Efficient maximum clique computation and enumeration over large sparse graphs*, VLDB J. **29** (2020), no. 5, 999–1022.
12. L. Chang, W. Li, and W. Zhang, *Computing a near-maximum independent set in linear time by reducing-peeling*, *Proc. of the 2017 ACM Intl. Conf. on Management of Data*, ACM, 2017, 1181–1196.
13. E. J. Cockayne, P. A. Dreyer Jr, S. M. Hedetniemi, and S. T. Hedetniemi, *Roman domination in graphs*, Discrete mathematics **278** (2004), no. 1-3, 11–22.
14. A. Conte, D. Firmani, M. Patrignani, and R. Torlone, *A meta-algorithm for finding large k-plexes*, Knowl. Inf. Syst. **63** (2021), no. 7, 1745–1769.
15. M. Cygan et al., *Parameterized algorithms*, vol. 4, Springer, 2015, .
16. T. A. Davis and Y. Hu, *The university of florida sparse matrix collection*, ACM Transactions on Mathematical Software (TOMS) **38** (2011), no. 1, 1–25.
17. Y. Ding, J. Z. Wang, and P. K. Srimani, *Self-stabilizing algorithm for maximal 2-packing with safe convergence in an arbitrary graph*, *2014 IEEE international parallel & distributed processing symposium workshops*, IEEE, 2014, 747–754.
18. E. D. Dolan and J. J. Moré, *Benchmarking optimization software with performance profiles*, Mathematical programming **91** (2002), no. 2, 201–213.
19. A. Flores-Lamas, J. A. Fernández-Zepeda, and J. A. Trejo-Sánchez, *Algorithm to find a maximum 2-packing set in a cactus*, Theoretical Computer Science **725** (2018), 31–51.
20. A. Flores-Lamas, J. A. Fernández-Zepeda, and J. A. Trejo-Sánchez, *A distributed algorithm for a maximal 2-packing set in halin graphs*, Journal of Parallel and Distributed Computing **142** (2020), 62–76.
21. M. Gairing, S. T. Hedetniemi, P. Kristiansen, and A. A. McRae, *Self-stabilizing algorithms for {k}-domination*, S.-T. Huang and T. Herman (eds.), *Self-Stabilizing Systems*, Springer Berlin Heidelberg, Berlin, Heidelberg, 2003, 49–60.
22. M. Gairing, W. Goddard, S. T. Hedetniemi, P. Kristiansen, and A. A. McRae, *Distance-two information in self-stabilizing algorithms*, Parallel Processing Letters **14** (2004), no. 03n04, 387–398.
23. M. Gairing, R. M. Geist, S. T. Hedetniemi, and P. Kristiansen, *A self-stabilizing algorithm for maximal 2-packing*, Nordic Journal of Computing **11** (2004), 1–11.
24. A. Gellner, S. Lamm, C. Schulz, D. Strash, and B. Zaválnij, *Boosting data reduction for the maximum weight independent set problem using increasing transformations*, *2021 Proceedings of the Workshop on Algorithm Engineering and Experiments (ALENEX)*, SIAM, 2021, 128–142.
25. J. Gramm, J. Guo, F. Hüffner, and R. Niedermeier, *Data reduction and exact algorithms for clique cover*, Journal of Experimental Algorithmics (JEA) **13** (2009), 2–2.
26. E. Großmann, K. Langedal, and C. Schulz, *Accelerating reductions using graph neural networks and a new concurrent local search for the maximum weight independent set problem* (2024).
27. E. Großmann, K. Langedal, and C. Schulz, *A Comprehensive Survey of Data Reduction Rules for the Maximum Weighted Independent Set Problem* (2024), .
28. E. Großmann, S. Lamm, C. Schulz, and D. Strash, *Finding Near-Optimal Weight Independent Sets at Scale*, Journal of Graph Algorithms and Applications **28** (2024), no. 1, 439–473.
29. J. Gu, W. Zheng, Y. Cai, and P. Peng, *Towards computing a near-maximum weighted independent set on massive graphs*, *Proceedings of the 27th ACM SIGKDD Conference on Knowledge Discovery & Data Mining*, 2021, 467–477.
30. J. Gómez Soto, J. Leaños, L. Ríos-Castro, and L. Rivera, *The packing number of the double vertex graph of the path graph*, Discrete Applied Mathematics **247** (2018), 327–340. URL https://www.sciencedirect.com/science/article/pii/S0166218X18301938.
31. W. Hale, *Frequency assignment: Theory and applications*, Proceedings of the IEEE **68** (1980), no. 12, 1497–1514.
32. M. M. Halldórsson, J. Kratochvíl, and J. A. Telle, *Independent sets with domination constraints*, K. G. Larsen, S. Skyum, and G. Winskel (eds.), *Automata, Languages and Programming, 25th International Colloquium, ICALP'98, Aalborg, Denmark, July 13-17, 1998, Proceedings*, *Lecture Notes in Computer Science*, vol. 1443, Springer, 1998, 176–187, . URL https://doi.org/10.1007/BFb0055051.
33. D. S. Hochbaum and D. B. S. R. work(s):, *A Best Possible Heuristic for the k-Center Problem*, Mathematics of Operations Research **10** (1985), no. 2, 180–184.
34. H. Jiang, D. Zhu, Z. Xie, S. Yao, and Z.-H. Fu, *A new upper bound based on vertex partitioning for the maximum k-plex problem*, Z.-H. Zhou (ed.), *Proceedings of the Thirtieth International Joint Conference on Artificial Intelligence, IJCAI-21*, International Joint Conferences on Artificial Intelligence Organization, 2021, 1689–1696, . Main Track.
35. I. Katsikarelis, M. Lampis, and V. Paschos, *Improved (In-)Approximability Bounds for d-Scattered Set*, Journal of Graph Algorithms and Applications **27** (2023), no. 3, 219–238.

36. S. Lamm, P. Sanders, C. Schulz, D. Strash, and R. F. Werneck, *Finding near-optimal independent sets at scale*, J. of Heuristics **23** (2017), no. 4, 207–229.
37. S. Lamm, C. Schulz, D. Strash, R. Williger, and H. Zhang, *Exactly solving the maximum weight independent set problem on large real-world graphs*, S. G. Kobourov and H. Meyerhenke (eds.), *Proceedings of the Twenty-First Workshop on Algorithm Engineering and Experiments, ALENEX 2019, San Diego, CA, USA, January 7-8, 2019*, SIAM, 2019, 144–158, . URL https://doi.org/10.1137/1.9781611975499.12.
38. K. Langedal, J. Langguth, F. Manne, and D. T. Schroeder, *Efficient minimum weight vertex cover heuristics using graph neural networks*, *20th International Symposium on Experimental Algorithms (SEA 2022)*, Schloss Dagstuhl-Leibniz-Zentrum für Informatik, 2022.
39. J. Leskovec and A. Krevl, *SNAP Datasets: Stanford large network dataset collection*, URL http://snap.stanford.edu/data (2014).
40. J. Lin, S. Cai, C. Luo, and K. Su, *A reduction based method for coloring very large graphs*, *Proceedings of the Twenty-Sixth International Joint Conference on Artificial Intelligence, IJCAI-17*, 2017, 517–523, .
41. F. Manne and M. Mjelde, *A memory efficient self-stabilizing algorithm for maximal k-packing*, *Stabilization, Safety, and Security of Distributed Systems: 8th International Symposium, SSS 2006, Dallas, TX, USA, November 17-19, 2006. Proceedings 8*, Springer, 2006, 428–439.
42. M. Mjedle, *k-packing and k-domination on tree graphs*, Master's thesis, University of Bergen, 2004.
43. B. C. S. Nogueira, R. G. S. Pinheiro, and A. Subramanian, *A hybrid iterated local search heuristic for the maximum weight independent set problem*, Optim. Lett. **12** (2018), no. 3, 567–583. URL https://doi.org/10.1007/s11590-017-1128-7.
44. P. V. Sander, D. Nehab, E. Chlamtac, and H. Hoppe, *Efficient traversal of mesh edges using adjacency primitives*, ACM Transactions on Graphics (TOG) **27** (2008), no. 5, 1–9.
45. P. Sanders, C. Schulz, and D. Wagner, *Benchmarking for graph clustering and partitioning*, Encyclopedia of social network analysis and mining Springer (2014).
46. Z. Shi, *A self-stabilizing algorithm to maximal 2-packing with improved complexity*, Information Processing Letters **112** (2012), no. 13, 525–531.
47. D. Strash, *On the power of simple reductions for the maximum independent set problem*, *Intl. Computing and Combinatorics Conf.*, Springer, 2016, 345–356.
48. D. Strash and L. Thompson, *Effective Data Reduction for the Vertex Clique Cover Problem*, SIAM, 2022. 41–53, .
49. J. A. Trejo-Sánchez, D. Fajardo-Delgado, and J. O. Gutierrez-Garcia, *A genetic algorithm for the maximum 2–packing set problem*, International Journal of Applied Mathematics and Computer Science **30** (2020), no. 1, 173–184.
50. J. A. Trejo-Sánchez and J. A. Fernández-Zepeda, *Distributed algorithm for the maximal 2-packing in geometric outerplanar graphs*, Journal of Parallel and Distributed Computing **74** (2014), no. 3, 2193–2202.
51. J. A. Trejo-Sánchez, J. A. Fernández-Zepeda, and J. C. Ramírez-Pacheco, *A self-stabilizing algorithm for a maximal 2-packing in a cactus graph under any scheduler*, International Journal of Foundations of Computer Science **28** (2017), no. 08, 1021–1045.
52. J. A. Trejo-Sánchez, F. A. Madera-Ramírez, J. A. Fernández-Zepeda, J. L. u. López-Martínez, and A. Flores-Lamas, *A fast approximation algorithm for the maximum 2-packing set problem on planar graphs*, Optimization Letters **14** (2023), 1435–1454.
53. J. A. Trejo-S'nchez and J. A. Fern'ndez-Zepeda, *A self-stabilizing algorithm for the maximal 2-packing in a cactus graph*, *2012 IEEE 26th international parallel and distributed processing symposium workshops & PhD Forum*, IEEE, 2012, 863–871.
54. V. Turau, *Efficient transformation of distance-2 self-stabilizing algorithms*, Journal of Parallel and Distributed Computing **72** (2012), no. 4, 603–612.
55. A. Verma, A. Buchanan, and S. Butenko, *Solving the maximum clique and vertex coloring problems on very large sparse networks*, INFORMS Journal on Computing **27** (2015), no. 1, 164–177.
56. C. Walshaw, *Graph partitioning archive*, URL https://chriswalshaw.co.uk/partition/ (2000).
57. K. Yamanaka, S. Kawaragi, and T. Hirayama, *Exact Exponential Algorithm for Distance-3 Independent Set Problem*, IEICE Transactions on Information and Systems **E102.D** (2019), no. 3, 499–501.

□

## APPENDIX

**TABLE** 1: Detailed graph properties. We present the number of vertices |V| and edges |E| of the instances as well as the average degree. The column weights shows the different weight assignements unit (u), uniform (uf), degree (d), geometric (g), hybrid (h) and from file (f).

| red2pack | n | m | weights | avg. deg |
|---|---|---|---|---|
| as-22july06 | 22 963 | 48 436 | u, uf, d, g, h | 4.2 |
| astro-ph | 16 706 | 121 251 | u, uf, d, g, h | 14.5 |
| cactus1000 | 1 000 | 1 008 | u, uf, d, g, h | 2.0 |
| caidaRouterLevel | 192 244 | 609 066 | u, uf, d, g, h | 6.3 |
| citationCiteseer | 268 495 | 1 156 647 | u, uf, d, g, h | 8.6 |
| coAuthorsCiteseer | 227 320 | 814 134 | u, uf, d, g, h | 7.2 |
| coAuthorsDBLP | 299 067 | 977 676 | u, uf, d, g, h | 6.5 |
| cond-mat-2003 | 31 163 | 120 029 | u, uf, d, g, h | 7.7 |
| cond-mat-2005 | 40 421 | 175 691 | u, uf, d, g, h | 8.7 |
| coPapersCiteseer | 434 102 | 16 036 720 | u, uf, d, g, h | 73.9 |
| coPapersDBLP | 540 486 | 15 245 729 | u, uf, d, g, h | 56.4 |
| hep-th | 8 361 | 15 751 | u, uf, d, g, h | 3.8 |
| loc-brightkite_edges | 56 739 | 212 945 | u, uf, d, g, h | 7.5 |
| loc-gowalla_edges | 196 591 | 950 327 | u, uf, d, g, h | 9.7 |
| netscience | 1 589 | 2 742 | u, uf, d, g, h | 3.5 |
| power | 4 941 | 6 594 | u, uf, d, g, h | 2.7 |
| road_central | 14 081 816 | 16 933 413 | u, uf, d, g, h | 2.4 |
| road_usa | 23 947 347 | 28 854 312 | u, uf, d, g, h | 2.4 |
| asia | 11 950 757 | 12 711 603 | uf, h | 2.1 |
| belgium | 1 441 295 | 1 549 970 | uf, h | 2.2 |
| europe | 50 912 018 | 54 054 660 | uf, h | 2.1 |
| germany | 11 548 845 | 12 369 181 | uf, h | 2.1 |
| great-britain | 7 733 822 | 8 156 517 | uf, h | 2.1 |
| italy | 6 686 493 | 7 013 978 | uf, h | 2.1 |
| luxembourg | 114 599 | 119 666 | uf, h | 2.1 |
| netherlands | 2 216 688 | 2 441 238 | uf, h | 2.2 |
| **fe** | **n** | **m** | **weights** | **avg. deg** |
| body | 45 087 | 163 734 | uf | 7.3 |
| ocean | 143 437 | 409 593 | uf | 5.7 |
| pwt | 36 519 | 144 794 | uf | 7.9 |
| rotor | 99 617 | 662 431 | uf | 13.3 |
| sphere | 16 386 | 49 152 | uf | 6.0 |
| **mesh** | **n** | **m** | **weights** | **avg. deg** |
| blob | 16 068 | 24 102 | uf | 3.0 |
| buddha | 1 087 716 | 1 631 574 | uf | 3.0 |
| bunny | 68 790 | 103 017 | uf | 3.0 |
| cow | 5 036 | 7 366 | uf | 2.9 |
| dragonsub | 600 000 | 900 000 | uf | 3.0 |
| dragon | 150 000 | 225 000 | uf | 3.0 |
| ecat | 684 496 | 1 026 744 | uf | 3.0 |
| face | 22 871 | 34 054 | uf | 3.0 |
| fandisk | 8 634 | 12 818 | uf | 3.0 |
| feline | 41 262 | 61 893 | uf | 3.0 |
| gameguy | 42 623 | 63 850 | uf | 3.0 |
| gargoyle | 20 000 | 30 000 | uf | 3.0 |
| turtle | 267 534 | 401 178 | uf | 3.0 |
| venus | 5 672 | 8 508 | uf | 3.0 |
| **osm** | **n** | **m** | **weights** | **avg. deg** |
| alabama-AM2 | 1 164 | 19 386 | f | 33.3 |
| alabama-AM3 | 3 504 | 309 664 | f | 176.7 |
| district-of-columbia-AM1 | 2 500 | 24 651 | f | 19.7 |
| district-of-columbia-AM2 | 13 597 | 1 609 795 | f | 236.8 |
| district-of-columbia-AM3 | 46 221 | 27 729 137 | f | 1 199.9 |
| florida-AM2 | 1 254 | 16 936 | f | 27.0 |
| florida-AM3 | 2 985 | 154 043 | f | 103.2 |
| georgia-AM3 | 1 680 | 74 126 | f | 88.2 |
| greenland-AM3 | 4 986 | 3 652 361 | f | 1 465.0 |

*Continued on next page*

Table 1 – *Continued from previous page*

| osm | *n* | *m* | weights | avg. deg |
|---|---|---|---|---|
| hawaii-AM2 | 2 875 | 265 158 | f | 184.5 |
| hawaii-AM3 | 28 006 | 49 444 921 | f | 3 531.0 |
| idaho-AM3 | 4 064 | 3 924 080 | f | 1 931.1 |
| kansas-AM3 | 2 732 | 806 912 | f | 590.7 |
| kentucky-AM2 | 2 453 | 643 428 | f | 524.6 |
| kentucky-AM3 | 19 095 | 59 533 630 | f | 6 235.5 |
| louisiana-AM3 | 1 162 | 37 077 | f | 63.8 |
| maryland-AM3 | 1 018 | 95 415 | f | 187.5 |
| massachusetts-AM2 | 1 339 | 35 449 | f | 52.9 |
| massachusetts-AM3 | 3 703 | 551 491 | f | 297.9 |
| mexico-AM3 | 1 096 | 47 131 | f | 86.0 |
| new-hampshire-AM3 | 1 107 | 18 021 | f | 32.6 |
| north-carolina-AM3 | 1 557 | 236 739 | f | 304.1 |
| oregon-AM2 | 1 325 | 57 517 | f | 86.8 |
| oregon-AM3 | 5 588 | 2 912 701 | f | 1 042.5 |
| pennsylvania-AM3 | 1 148 | 26 464 | f | 46.1 |
| rhode-island-AM2 | 2 866 | 295 488 | f | 206.2 |
| rhode-island-AM3 | 15 124 | 12 622 219 | f | 1 669.2 |
| utah-AM3 | 1 339 | 42 872 | f | 64.0 |
| vermont-AM3 | 3 436 | 1 136 164 | f | 661.3 |
| virginia-AM2 | 2 279 | 60 040 | f | 52.7 |
| virginia-AM3 | 6 185 | 665 903 | f | 215.3 |
| washington-AM2 | 3 025 | 152 449 | f | 100.8 |
| washington-AM3 | 10 022 | 2 346 213 | f | 468.2 |
| west-virginia-AM3 | 1 185 | 125 620 | f | 212.0 |
| snap | *n* | *m* | weights | avg. deg |
| as-skitter | 1 696 415 | 11 095 298 | uf | 13.1 |
| ca-AstroPh | 18 772 | 198 050 | uf | 21.1 |
| ca-CondMat | 23 133 | 93 439 | uf,f | 8.1 |
| ca-GrQc | 5 241 | 14 484 | uf,f | 5.5 |
| ca-HepPh | 12 008 | 118 489 | uf | 19.7 |
| ca-HepTh | 9 877 | 25 973 | uf | 5.3 |
| com-amazon | 334 863 | 925 869 | f | 5.5 |
| com-youtube | 1 134 890 | 2 987 624 | f | 5.3 |
| email-Enron | 36 692 | 183 831 | uf,f | 10.0 |
| email-EuAll | 265 214 | 364 481 | uf | 2.7 |
| loc-gowalla_edges | 196 591 | 950 327 | f | 9.7 |
| p2p-Gnutella04 | 10 876 | 39 994 | uf | 7.4 |
| p2p-Gnutella05 | 8 846 | 31 839 | uf | 7.2 |
| p2p-Gnutella06 | 8 717 | 31 525 | uf | 7.2 |
| p2p-Gnutella08 | 6 301 | 20 777 | uf | 6.6 |
| p2p-Gnutella09 | 8 114 | 26 013 | uf | 6.4 |
| p2p-Gnutella24 | 26 518 | 65 369 | uf | 4.9 |
| p2p-Gnutella25 | 22 687 | 54 705 | uf | 4.8 |
| p2p-Gnutella30 | 36 682 | 88 328 | uf | 4.8 |
| p2p-Gnutella31 | 62 586 | 147 892 | uf | 4.7 |
| roadNet-CA | 1 965 206 | 2 766 607 | uf | 2.8 |
| roadNet-PA | 1 088 092 | 1 541 898 | uf,f | 2.8 |
| roadNet-TX | 1 379 917 | 1 921 660 | uf | 2.8 |
| soc-Epinions1 | 75 879 | 405 740 | uf | 10.7 |
| soc-LiveJournal1 | 4 847 571 | 42 851 237 | uf | 17.7 |
| soc-pokec-relationships | 1 632 803 | 22 301 964 | uf | 27.3 |
| soc-Slashdot0811 | 77 360 | 469 180 | uf | 12.1 |
| soc-Slashdot0902 | 82 168 | 504 230 | uf | 12.3 |
| web-BerkStan | 685 230 | 6 649 470 | uf,f | 19.4 |
| web-Google | 875 713 | 4 322 051 | uf | 9.9 |
| web-NotreDame | 325 729 | 1 090 108 | uf,f | 6.7 |
| web-Stanford | 281 903 | 1 992 636 | uf | 14.1 |
| wiki-Talk | 2 394 385 | 4 659 565 | uf | 3.9 |
| wiki-Vote | 7 115 | 100 762 | uf | 28.3 |
| ssmc | *n* | *m* | weights | avg. deg |
| ca2010 | 710 145 | 1 744 683 | f | 4.9 |
| fl2010 | 484 481 | 1 173 147 | f | 4.8 |
| ga2010 | 291 086 | 709 028 | f | 4.9 |
| il2010 | 451 554 | 1 082 232 | f | 4.8 |
| nh2010 | 48 837 | 117 275 | f | 4.8 |
| ri2010 | 25 181 | 62 875 | f | 5.0 |